\documentclass[]{jfm}

\usepackage{graphicx}
\usepackage{newtxtext}
\usepackage{newtxmath}
\usepackage{natbib}
\usepackage{hyperref}
\usepackage{bm}
\hypersetup{
    colorlinks = true,
    urlcolor   = blue,
    citecolor  = black,
}

\newcommand{\RomanNumeralCaps}[1]
\linenumbers

\newcommand{\vx}{\bm{x}}
\newcommand{\vpbar}{\bar{\bar{\bm{\varphi}}}}
\newcommand{\vpstar}{{\bm{\varphi}}^*_p}
\newcommand{\vpbarp}{\bar{\bar{\bm{\varphi}}}_p}

\newcommand{\vp}{\bm{\varphi}}

\newcommand{\vXi}{\bm{\Xi}}
\newcommand{\vXia}[1]{\bm{\Xi}^{#1}}

\newcommand{\vXip}[1]{\bm{\Xi}^{#1}_p}
\newcommand{\vxip}[1]{\bm{\xi}^{#1}_p}
\newcommand{\va}{\bm{a}}
\newcommand{\vu}{\bm{u}}
\newcommand{\barL}[1]{\overline{#1}^\mathrm{L}}
\def\d{\mathrm{d}}
\def\e{\mathrm{e}}
\def\i{\mathrm{i}}

\usepackage{xcolor}

\title{Lagrangian filtering for wave--mean flow decomposition}

\author{Lois E. Baker\aff{1}
  \corresp{\email{lois.baker@ed.ac.uk}},
  Hossein A. Kafiabad\aff{2}, 
  Cai Maitland-Davies\aff{2}
 \and Jacques Vanneste\aff{1}}

\affiliation{\aff{1}School of Mathematics, University of Edinburgh, Edinburgh, UK
\aff{2}Department of Mathematical Sciences, Durham University, Durham, UK}

\begin{document}
\maketitle

\begin{abstract} 

Geophysical flows are typically composed of wave and mean motions with a wide range of overlapping temporal scales, making separation between the two types of motion in wave-resolving numerical simulations challenging. Lagrangian filtering -- whereby a temporal filter is applied in the frame of the flow -- is an effective way to overcome this challenge, allowing clean separation of waves from mean flow based on frequency separation in a Lagrangian frame. Previous implementations of Lagrangian filtering have used particle tracking approaches, which are subject to large memory requirements or difficulties with particle clustering. \citet[][KV23]{kafiabadComputingLagrangianMeans2023} recently proposed a novel method for finding Lagrangian means without particle tracking by solving a set of partial differential equations alongside the governing equations of the flow. In this work, we adapt the approach of KV23 to develop a flexible, on-the-fly, PDE-based method for Lagrangian filtering using arbitrary convolutional filters. We present several different wave--mean decompositions, demonstrating that our Lagrangian methods are capable of recovering a clean wave-field from a nonlinear simulation of geostrophic turbulence interacting with Poincar{\'e} waves. 

\end{abstract}

\begin{keywords}

\end{keywords}


\section{Introduction}
\label{sec:Intro}

The motions of the ocean and atmosphere involve processes with a wide range of spatial and temporal scales. In particular, fast internal waves propagate throughout these stratified, rotating fluids, interacting with slower eddies and currents. Internal waves play a key role in the momentum and energy budgets of the ocean and atmosphere, forcing the mean flow, transferring energy between large and small scales, and causing turbulent mixing when they break \citep{NaveiraGarabato2013,Waterhouse2014,MacKinnon2017,Whalen2018,Shakespeare2019,Whalen2020}. Understanding and quantifying the role of internal waves in shaping the larger scale circulation is key to designing accurate climate models, since their spatial scales cannot be directly resolved and must instead be parameterised. Wave-resolving numerical simulations are an important tool for understanding the physics that underpins these parameterisations.

A primary challenge in the study of interactions of internal waves with the non-wave flow in numerical wave-resolving simulations is separating the processes so that they can be quantified and their physics understood. We hereafter refer to the non-wave flow as the mean flow, with the understanding that this does not place any restriction on the temporal or spatial scales of the mean flow at this stage. Many different methods have been proposed for separating waves from the mean flow, often relying on defining a `balanced' mean flow based on dynamical considerations (for example, geostrophic balance in the limit of small Rossby number, or its higher order variants \citep{vallisAtmosphericOceanicFluid2017,Vanneste2013}), and taking the wave component of the flow to be the `unbalanced' residual \citep{Buhler2014book}. However, recent studies have highlighted the importance of mean flow regimes that are $O(1)$ in the Rossby number, indicating that inertial and rotational forces on the flow are comparable, and may therefore be `unbalanced' \citep{McWilliams2016a,Taylor2022}. These flows are termed submesoscale in the ocean, and mesoscale in the atmosphere. 

Submesoscale currents and internal waves are both typically energetic in the important surface and bottom boundary layers of the ocean, and despite our inability to directly capture their effects in climate models, they are starting to be regularly resolved in realistic, regional high-resolution numerical models \citep[e.g.][]{Nagai2015,Bachman2017,Su2018,bakerBoundaryUpwellingAntarctic2023}. Understanding the physics of their interactions (to ultimately inform parameterisations in coarser models) has therefore become a topic of significant recent interest \citep[e.g.][]{tedescoSpatiotemporalCharacteristicsNearSurface2023, barkanEddyInternalWave2024,thomasBlockedDrainpipesSmoking2024}, requiring an effective way to separate the wave-like part of the flow from other motions. In particular, to study wave generation, propagation, and mixing it is important to know the wave-like part of the flow, not only the mean flow. This task is more challenging in practice than only finding the mean flow, since the waves are often lower amplitude than the mean flow and therefore more easily contaminated by imperfect decompositions. 

Averaging techniques based on spatial or temporal scales are often used to define a mean state, with the wave component of the flow defined as the perturbation from this mean. In particular, weighted averages can be used to control the temporal frequencies or spatial wavenumbers that constitute the mean state (see \S \ref{sec:freqfilters}). We refer to these weighted averages as filters, but keep in mind that filtering is just a special case of averaging.  

Filtering on wavenumber or frequency also presents problems for separating waves from the mean flow. Internal waves can have wavelengths ranging from hundreds of metres to hundreds of kilometres, often overlapping in spatial scales with motions such as oceanic submesoscales. Moreover, whilst internal waves are often considered `fast' and the non-wave flow `slow', this is a simplification. Internal waves in geophysical flows have an intrinsic frequency greater than $f$, where $f$ is the Coriolis parameter and quantifies the rate of Earth's rotation \citep[although this can be modified to an `effective' Coriolis parameter by background vorticity or baroclinicity of the flow;][]{Kunze1985,whittNearInertialWavesStrongly2013}. However, the intrinsic frequency is the frequency in the frame of the flow, rather than the rest frame. Due to Doppler shifting of internal waves by the mean flow, there is no such frequency constraint on internal waves in the rest frame. Indeed, an important class of internal waves in the ocean and atmosphere is steady, topographically-generated lee waves, which have zero frequency in the fluid's rest frame. 

Furthermore, when the waves have non-negligible amplitude they perturb the mean flow with the wave frequency, so that the non-wave component of a flow found by temporally filtering in the rest frame is `blurred' by the presence of waves (KV23). These two effects often render temporal filtering in the rest frame (Eulerian temporal filtering) problematic.

A proposed solution to these filtering difficulties is to perform the temporal filter not in the rest frame, but in the frame moving with the flow. This has been termed \textit{Lagrangian filtering} \citep{Nagai2015,Shakespeare2017c}. The key assumption in the geophysical context is that internal waves can be defined by a super-inertial ($> f$) intrinsic frequency \citep{polzinRegionalCharacterizationsOceanic2011}, although $f$ can be replaced by an effective inertial frequency when background vorticity is strong \citep{Kunze1985,Rama2022}. Whilst Lagrangian filtering in numerical simulations is a relatively recent development, the power of Lagrangian mean theories has been well-known since pioneering theoretical work of \citet{brethertonGeneralLinearizedTheory1971}, \citet{sowardKinematicTheoryLarge1972}, and \citet{andrewsExactTheoryNonlinear1978}. In particular, the development of Generalised Lagrangian Mean (GLM) theory by \cite{andrewsExactTheoryNonlinear1978} showed that a tractable definition of Lagrangian means is available and, when applied to the equations of fluid motion, leads to simple and natural equations for the Lagrangian mean that are not available for the Eulerian mean \citep{andrewsExactTheoryNonlinear1978,Buhler2014book,gilbertGeometricGeneralisedLagrangianmean2018,kafiabadWaveaveragedBalanceSimple2021,gilbertGeometricApproaches2024}.  

Despite the many benefits of Lagrangian temporal averaging (and, as a special case, Lagrangian filtering) over Eulerian averaging, it is not typically used in the analysis of numerical simulations due to computational challenges: numerical simulations are usually Eulerian in nature, with data being defined at fixed spatial grid points. Previous approaches have used particle tracking methods, whereby synthetic passive particles are seeded in a numerical simulation and advected by the flow velocities, either during the simulation or afterwards using saved data. The scalar fields recorded at the particle locations can then be used to formulate Lagrangian averages \citep{Nagai2015,Shakespeare2017c,Shakespeare2018,Shakespeare2019,bachmanParticleBasedLagrangianFiltering2020}. Various difficulties with this approach include the computational expense of advecting particles and the potential of particles to cluster in certain parts of the domain, making a domain-wide Lagrangian average subject to potentially inaccurate interpolations. A recent open-source Python package developed by \cite{Shakespeare2021a} overcomes the latter of these difficulties by performing the particle tracking on offline simulation data and seeding particles at the midpoint of the interval of interest, from which they are tracked back and forth. This method has been successfully used to filter internal waves in a number of studies, but relies on large amounts of high spatial and temporal resolution simulation data being saved and processed \citep[e.g.][]{Shakespeare2021a,BakerMashayek2022,tedescoSpatiotemporalCharacteristicsNearSurface2023,jonesUsingLagrangianFiltering2023}. 

An alternative method for finding the Lagrangian mean was recently proposed by \citet[][hereafter KV23]{kafiabadComputingLagrangianMeans2023}, following on from a previous grid-based method for the same procedure \citep{kafiabadGridbasedCalculationLagrangian2022}. They showed that it is possible to define partial Lagrangian mean fields that lead to a set of partial differential equations (PDEs) that only depend on the current simulation time. These PDEs can be solved alongside the governing equations of the flow over the averaging interval, after which the final value of the partial Lagrangian means is equal to the full Lagrangian mean of interest. For each interval over which the Lagrangian mean equations are solved, only one instance of the full Lagrangian mean is computed, so the Lagrangian mean can either be found at a coarse temporal resolution, or multiple sets of the PDEs can be solved simultaneously to achieve a higher temporal resolution. This method allows the Lagrangian mean to be found on-the-fly, with no expensive data writing, storage or post-processing required. The Lagrangian mean equations can be solved with the same scheme as the governing equations of the flow. 

The Lagrangian mean found using the KV23 approach is the special case of an unweighted mean over a finite interval, often referred to as a `top-hat' mean. However, in order to control the frequencies that are retained by the mean field, i.e. to apply a Lagrangian frequency filter, a weighted mean is needed. In this work, we extend the method of KV23 to one for a general convolutional weighted mean. This allows us to perform Lagrangian filtering on-the-fly in a numerical simulation without particle tracking. We present three strategies for this purpose -- two of which were previously presented by KV23, and a third, new strategy that avoids some issues associated with the other two. 

Once the Lagrangian mean fields are computed, it is possible to define the corresponding perturbation fields in a number of ways. We therefore also carefully present several different Lagrangian wave--mean decompositions and their properties. Whilst the motivation here is to enable identification of internal inertia-gravity waves, this method is flexible in that any intrinsic frequency criterion can be used to define the wave-like perturbations. Although we focus on geophysical flows, our method can be used for any multi-time-scale flows, such as those in astrophysical or biological fluids.

This paper is structured as follows. In \S \ref{sec:mean_formulation}, we introduce the weighted Lagrangian mean, some of its important properties, and consider desirable forms of the weight function. In \S \ref{sec:formulation}, we derive the on-the-fly method for solving for the Lagrangian mean. In \S \ref{sec:numericalmodel} we introduce a rotating shallow water model that we use as a test-bed for the Lagrangian mean computation, and in \S \ref{sec:meanresults} we show results of solving the Lagrangian mean equations alongside this model for the various strategies. Then, in \S \ref{sec:wavemeandecomp} we return to a more theoretical look at how a wave--mean decomposition should be defined, before presenting results of these wave--mean decompositions in the shallow water model in \S \ref{sec:waveresults}. We discuss potential errors in \S \ref{sec:interpolation}, and our methods and results in \S \ref{sec:discussion}.

\section{Lagrangian mean formulation}\label{sec:mean_formulation}
\subsection{Weighted averages and frequency filters}\label{sec:freqfilters}
We begin by defining a standard weighted time average at the reference time $t^*$ of some scalar quantity $g(t)$ as
\begin{equation}\label{timeav}
    \bar{g}(t^*) = \int_{-\infty}^{\infty} g(s)F(s,t^*) \,\d s\,.
\end{equation}

We design a weight function that acts as a frequency filter on $g$ at time $t^*$. Consider the Fourier Transform of $g$, given by
\begin{equation}\label{frequencyspectrum}
 \hat{g}(\omega) = \int_{-\infty}^{\infty}g(s)\e^{- \i \omega s}\,\d s\,,  
\end{equation}
then the frequency filtered scalar $\bar{g}$ at time $t^*$ is given by
\begin{equation}\label{gfiltered}
    \bar{g}(t^*) = \frac{1}{2\pi}\int_{-\infty}^{\infty} \hat{G}(\omega)\hat{g}(\omega)\e^{ \i \omega t^*} \, \d\omega\,,
\end{equation}
where $\hat{G}(\omega)$ weights certain frequencies -- for example, when $\hat{G}(\omega) = 1$ on $[-\omega_c,\omega_c]$ and is zero otherwise, $\bar{g}(t^*)$ is low-pass filtered with cut-off frequency $\omega_c$. 

Writing \eqref{gfiltered} in the form of \eqref{timeav} gives
\begin{equation}
    \bar{g}(t^*) =\int_{-\infty}^{\infty}g(s)G(t^*-s)\,\d  s\,,
\end{equation}
where
\begin{equation}\label{FTF}
G(t) = \frac{1}{2\pi}\int_{-\infty}^\infty \hat{G}(\omega) \e^{\i \omega t}\, \d \omega   \,,
\end{equation}
thus using the convolutional weight function $F(s,t^*) = G(t^* - s)$ in \eqref{timeav} gives the frequency filter of $g$ at time $t^*$. $G(t)$ can also be described as an impulse response, and $\hat{G}(\omega)$ as the corresponding frequency response of the filter. If $G_{\mathrm{LP}}$ describes a low-pass filter with cut-off $\omega_c$ described above, then
\begin{equation}\label{lowpass}
    G_{\mathrm{LP}}(t) = \frac{\sin(\omega_c t)}{\pi t}\,,
\end{equation}
and the top-hat mean over an interval of length $2T$ is given by
\begin{equation}\label{tophat}
    G_{\mathrm{TH}}(t) = (H(t+T) - H(t-T))/2T\,,
\end{equation}
where $H(\cdot)$ is the Heaviside step function.
We hereafter consider convolutional weight functions of the form $F(s,t^*) = G(t^* - s)$.
\subsection{Lagrangian averaging}
If a time average is calculated at a fixed point in space, then it is an \textit{Eulerian} time average. If it is instead calculated along the trajectory of a particle travelling with the fluid velocity $\vu(\vx,t)$, it is a \textit{Lagrangian} time average. 

We define a flow map $\vp(\va,t)$, which gives the position of a particle labelled by $\va$ at time $t$. The label $\va$ could be taken to be the position of the particle at time $t=0$, so that $\vp(\va,0) = \va$, and in general, we think of $\vp, \va \in \mathbb{R}^2$ or $\mathbb{R}^3$. Following KV23, we then define the weighted Lagrangian mean flow map as
\begin{equation}\label{phibardef}
\vpbar(\va ,t^*) = \int_{-\infty}^{\infty} G(t^*-s)\vp(\va,s)\,\d s\,,
\end{equation}
where $t^*$ is the time to which the Lagrangian mean is assigned. Following \cite{gilbertGeometricApproaches2024} we use the double-bar notation to avoid confusion with the straightforward Eulerian average.
We then define the weighted generalised Lagrangian mean of a scalar field $f(\vx,t)$ by
\begin{equation}\label{fbardef}
\barL{f}(\vpbar(\va,t^*),t^*) = \int_{-\infty}^{\infty} G(t^*-s)f(\vp(\va,s),s) \, \d s\,.
\end{equation}
For comparison, we also define the corresponding Eulerian mean
\begin{equation}\label{fbarEdef}
\overline{f}^\mathrm{E}(\vx,t^*) = \int_{-\infty}^{\infty} G(t^*-s)f(\vx,s) \, \d s\,.
\end{equation}

\subsection{Defining a valid weight function}\label{sec:definingmean}
We require that the weight function $G(t^*-s)$ satisfies the normalisation condition
\begin{equation}\label{normalisation}
    \int_{-\infty}^{\infty} G(t^*-s)\,\d  s = 1\,.
\end{equation}
From \eqref{phibardef}, this is equivalent to requiring that the mean position of a stationary particle (i.e. when flow velocity $\vu = 0$ so that the flow map $\vp$ is independent of time) is its own position. Without this natural assumption, our methods for finding $\barL{f}$ in a periodic domain break down (see discussion of suitable domains and boundary conditions in \S \ref{sec:BCs}). 

Writing $G$ in terms of its Fourier transform $\hat{G}$ (defined in \eqref{FTF}) shows that \eqref{normalisation} is equivalent to $\hat{G}(0) = 1$. The filter must therefore include the zero frequency. The filters described in equations \eqref{lowpass} and \eqref{tophat} satisfy this criterion, but a high pass filter, for example, could not be used to define a mean flow. However, we could define a weight function that removes some frequencies $\omega$ in a specified interval $0 < |\omega_1| < |\omega| < |\omega_2|$ by setting:
\begin{equation}\label{lowbandpassfilter}
    \hat{G}(\omega) = \begin{cases}
        0 , \hspace{1cm} |\omega_1| < |\omega| < |\omega_2| \\
        1 , \hspace{1cm} \mathrm{otherwise}\,.
    \end{cases}
\end{equation}
We later show an example of this filter in figure \ref{fig:lowbandpass}.

We would also like the weight function to be such that $\barL{f}$ (defined in \eqref{fbardef}) satisfies a property that we expect of a mean, namely that the mean is unchanged by reapplying the averaging operation:
\begin{equation}\label{meanofmean}
    \overline{\,\barL{f}}^{\,\mathrm{L}} (\vx,t^*)\, \overset{?}{=} \barL{f} (\vx,t^*)\,.
\end{equation}
Since the Lagrangian mean of a scalar depends on the flow with which it is advected, there is some ambiguity in the notation on the $LHS$ of equation \eqref{meanofmean}. We formalise this statement in Appendix \ref{app:meanofmean}, and note here that the Lagrangian mean of $\barL{f}$ is taken with respect to the Lagrangian mean flow defined by the map $\vpbar$. 

Equation \eqref{meanofmean} is satisfied only when $\hat{G}(\omega) = 0$ or $1$, or some piece-wise combination of each. A proof of this is given in Appendix \ref{app:meanofmean}. This is equivalent to requiring that $\hat{G}$ represents a perfect band-pass filter. In this case, the Lagrangian filtered field $\barL{f}$ behaves as we hope a `mean' field should, in that it contains no (Lagrangian) high frequencies. In practice, filters rarely exactly satisfy the condition \eqref{meanofmean}. Perfect band-pass filters tend to suffer from spectral ringing at the cut-off frequency, so other imperfect filters such as the Butterworth filter are often used instead \citep{Rama2022}. Here, the wave frequency considered in the numerical model in \S \ref{sec:numericalmodel} is not close to the cut-off frequency, so ringing is not an issue. We therefore only consider perfect band-pass filters so that the mean field is not expected to contain any wave signal. Using a 4th order Butterworth filter gives indistinguishable results in our case, although it does allow shorter averaging intervals (see discussion of figure \ref{fig:strat13_comparison}), which makes it preferable in practice, even though \eqref{meanofmean} is not perfectly satisfied.

We note here that there is no assumption of time-scale separation between the `slow' mean flow and the `fast' motions to be filtered. The original formulation of GLM theory by \cite{andrewsExactTheoryNonlinear1978} defines the Lagrangian average in an abstract way to apply to ensemble averages. To apply it to temporal averages such as the one we compute requires an assumption that the mean flow is `frozen' during the averaging operation \citep{Buhler2014book}, and this is explained further with an example illustrating the difference between the formulations in Appendix \ref{app:timescales}. 

\subsection{Lagrangian mean velocity}\label{sec:Lagrangianvel}
By definition, the flow map $\vp$ satisfies
\begin{equation}\label{udef}
    \vu(\vp(\va,t),t) = \frac{\partial \vp}{\partial t}(\va,t)\,,
\end{equation}
where $\vu$ is the fluid velocity. The Lagrangian mean velocity $\bar{\bar{\vu}}$ is then defined to be the velocity of a particle moving along a Lagrangian mean trajectory, that is,
\begin{equation}\label{LMUdef}
    \bar{\bar{\vu}}(\vpbar(\va,t^*),t^*) = \frac{\partial \vpbar}{\partial t^*}(\va,t^*)\,.
\end{equation}
However, another velocity $\barL{\vu}$ can be defined by taking the Lagrangian mean of each component of the velocity $\vu$ treated as scalars (see \citet{gilbertGeometricGeneralisedLagrangianmean2018} and \citet{gilbertGeometricApproaches2024} for other, more geometric definitions of $\barL{\vu}$). The averaging operation is such that $\bar{\bar{\vu}} = \barL{\vu}$ for the class of convolutional weight functions considered here. This is a special case of the more general result
\begin{equation}\label{materialderivativerel}
    \bar{\bar{D}}\barL{f} = \overline {\,D  f\,}^{\mathrm{L}}\,,
\end{equation}
where
\begin{align}\label{materialderivativedef}
    D &\equiv \frac{\partial}{\partial t} + \vu\bcdot \nabla \,, \\
    \bar{\bar{D}} &\equiv \frac{\partial}{\partial t^*} + \bar{\bar{\vu}}\bcdot \nabla \,.
\end{align}
This result is shown in Appendix \ref{app:materialderivative}, with the result $\bar{\bar{\vu}} = \barL{\vu}$ found by considering $f(\vx,t^*) = \vx$. Equation \eqref{materialderivativerel} is one of the most powerful results of the Lagrangian formalism -- it means that material conservation laws and scalar transport relations are inherited naturally by the corresponding Lagrangian means \citep{andrewsExactTheoryNonlinear1978}. 

We now formulate a PDE-based method for calculating these Lagrangian mean quantities, extending the method of KV23 to include a general convolutional weight function $G(t^*-s)$, which allows us to use a Lagrangian filter for wave--mean decomposition. 

\section{Formulation of on-the-fly method}\label{sec:formulation}
KV23 developed a method for finding the top-hat Lagrangian mean $\barL{f}$ of a scalar field $f$, as defined in \eqref{fbardef} (with $G(t) = (H(t+T) - H(t-T))/2T$), by formulating equations for the `partial Lagrangian means' and evolving them in a numerical simulation alongside the governing equations of the flow. Here, we re-derive this method for a weighted Lagrangian mean, although we have the specific application of a low-pass filter in mind. 

KV23 presented two strategies for finding $\barL{f}$. Strategy 1 solves first for an auxiliary mean function, before using a remapping to recover $\barL{f}$, whereas strategy 2 solves directly for $\barL{f}$. Both of these strategies have particular advantages and disadvantages, which we discuss further later. Here, we rederive these two strategies for a weighted mean, and present a new third strategy that circumvents some difficulties with strategies 1 and 2. 
\begin{figure}
    \includegraphics[width=\textwidth]{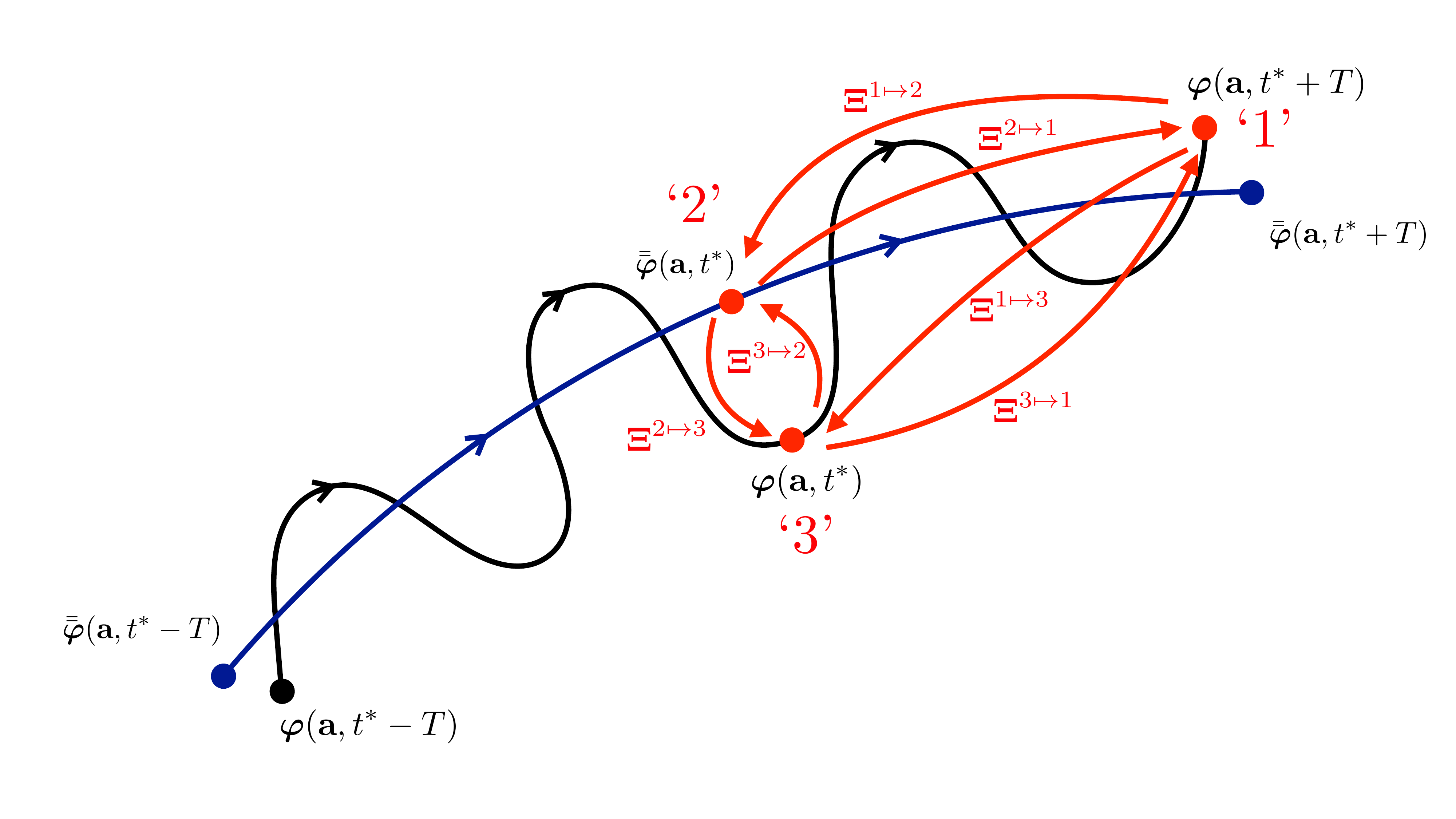}
    \caption{Schematic of a particle trajectory (black) with label $\va$ in the interval $[t^*-T,t^* + T]$, with positions labelled by the flow map $\vp(\va,t)$. The mean particle trajectory on the same interval is shown in blue, with positions labelled by the mean flow map $\vpbar(\va,t^*)$. Red arrows indicate the maps $\vXia{i \mapsto j}$ from position $i$ to position $j$, where position 1 is the trajectory endpoint $\vp(\va,t^* +T)$, position 2 is the trajectory mean $\vpbar(\va,t^*)$, and position 3 is the trajectory midpoint $\vp(\va,t^*)$. }
    \label{fig:trajectorydiagram}
\end{figure}

\subsection{Definition of full Lagrangian means}\label{sec:fullLagmean}
We now approximate the average over an infinitely long interval in \eqref{phibardef}-\eqref{fbardef} by one over a finite interval $[t^*-T,t^*+T]$, which is centred on the time $t^*$ at which the average is defined. This need not be the case, but it is the natural choice for even weight functions $G$, which correspond to real frequency response functions $\hat{G}$ (see \eqref{FTF}).

The mean flow map is now defined by (c.f. \eqref{phibardef})
\begin{equation}\label{vpbardef}
    \vpbar(\va,t^*) = \int_{t^*-T}^{t^*+T} G(t^*-s) \vp(\va,s)\,\d s\,,
\end{equation}
where $G(t^*-s)$ satisfies the normalisation \eqref{normalisation} over the interval $[t^*-T,t^*+T]$. 

We introduce some notation defining rearrangements of the mean scalar $\barL{f}$ that depend on different spatial coordinates:
\begin{equation}
    \barL{f}(\vpbar(\va,t^*),t^*) = \tilde f(\vp(\va,t^*+T),t^*) = f^*(\vp(\va,t^*),t^*) = \int_{t^*-T}^{t^*+T} G(t^*-s) f(\vp(\va,s),s)\,\d s \,.\label{fbardefs}
\end{equation}
$\barL f$, $\tilde f$ and $f^*$ all encode the Lagrangian mean of $f$, but they use different independent variables to do so: $\barL f(\vx,t^*)$ is the Lagrangian mean for the particle whose mean position is $\vx$,  $\tilde f(\vx,t^*)$ is the Lagrangian mean for the particle whose position at time $t^* + T$ is $\vx$, and $f^*(\vx,t^*)$ is the Lagrangian mean for the particle whose position at time $t^*$ is $\vx$. The three functions are rearrangements of each other, that is, related  via composition with (not necessarily volume-preserving) maps. 

Despite simply being rearrangements of each other, each of the definitions in \eqref{fbardefs} will be useful to us. $\barL f$ is the generalised Lagrangian mean that we are looking to find, and is the true Lagrangian mean in that it satisfies properties such as \eqref{meanofmean} and \eqref{materialderivativerel}. $\tilde{f}$ and $f^*$ are auxiliary fields that will help us to derive $\barL{f}$ in two of our strategies, and we will also demonstrate that $f^*$ is useful in itself to extract the wave field. Hence, one may want to compute it in addition to $\barL {f}$. 

We also need to define the set of maps $\vXi$ between each of the spatial independent variables that effect the rearrangements of $\barL{f}$ in \eqref{fbardefs}. For this purpose, the label `1' refers to the trajectory endpoint position $\vp(\va,t^*+T)$, `2' refers to the trajectory mean position $\vpbar(\va,t^*)$, and `3' refers to the trajectory midpoint position $\vp(\va,t^*)$. We use this convention because strategy `$i$' directly finds the Lagrangian mean in \eqref{fbardefs} with spatial independent variable `$i$', for $i \in \{1,2,3\}$. For example, strategy 1 solves directly for $\tilde{f}$. We define the map $\vXia{i\mapsto j}$ to map from the $i$ coordinate to the $j$ coordinate, such that $\vXia{i\mapsto j}$ is the identity map for $i=j$, $\left(\vXia{i\mapsto j}\right)^{-1} = \vXia{j\mapsto i}$ and
\begin{align}
\vXia{1\mapsto 2}(\vp(\va,t^*+T),t^*) &= \vpbar(\va,t^*)\,, \label{Xifullmaps1}\\
\vXia{1\mapsto 3}(\vp(\va,t^*+T),t^*) &= \vp(\va,t^*)\,,\label{Xifullmaps2}\\
\vXia{3\mapsto 2}(\vp(\va,t^*),t^*) &= \vpbar(\va,t^*)\,.\label{Xifullmaps3}
\end{align}
These maps are illustrated in the schematic in figure \ref{fig:trajectorydiagram}.

\subsection{Definition of partial Lagrangian means}

As in KV23, we now define a corresponding set of `partial' Lagrangian mean fields, that is, fields obtained by carrying out the averaging integration from $t^* - T$ to some $t < t^* + T$. 
By finding PDEs for these partial fields and evolving them over the averaging interval alongside the governing equations for the flow, the full Lagrangian mean fields in \S \ref{sec:fullLagmean} are obtained when $t = t^* + T$. The subscript $p$ always denotes a `partial' field, and these fields evolve with time $t$, while the time $t^*$ at which the Lagrangian mean is assigned is a fixed parameter. In the definitions of the partial fields, we drop the dependence on $t^*$ for readability, since everything in this section refers to one fixed averaging time $t^*$. 

First, we define a partial mean flow map to correspond to \eqref{vpbardef}, namely
\begin{equation}\label{vpbarpdef}
    \vpbarp(\va,t) = \int_{t^*-T}^t G(t^*-s) \vp(\va,s)\,\d s + \vp(\va,t)\left(1 - \int_{t^*-T}^t G(t^*-s)\,\d s\right)\,,
\end{equation}
so that $\vpbarp(\va,t^*+T) = \vpbar(\va,t^*)$. This particular form of $\vpbarp$ (in particular the second term which vanishes when $t = t^* + T$) is needed for a similar reason to that discussed in \S \ref{sec:definingmean}, namely that the partial mean position of a stationary particle should be the position itself, so that the image of the partial mean flow map is the same as that of the flow map itself. 

We then define the partial equivalents of \eqref{fbardefs}:
\begin{equation}
    \barL{f}_p(\vpbarp(\va,t),t) = \tilde f_p(\vp(\va,t),t) = f^*_p(\vpstar(\va,t),t) = \int_{t^*-T}^t G(t^*-s) f(\vp(\va,s),s)\,\d s\,, \label{fbarpdefs}
\end{equation}
where a second term corresponding to that used in \eqref{vpbarpdef} is not necessary here and is omitted for convenience. The new partial coordinate corresponding to $\vp(\va,t^*)$ is $\vpstar(\va,t)$, given by
\begin{equation}\label{vpstar}
    \vpstar(\va,t) = 
    \begin{cases}
    \vp(\va,t), \hspace{1cm} t < t^* \\
    \vp(\va,t^*), \hspace{1cm} t \geq t^*\,.
    \end{cases}
\end{equation}
This form of $\vpstar$ is necessary because using $\vp(\va,t^*)$ as the coordinate from the beginning would violate causality. Other definitions, such as $\vpstar(\va,t) = \vp(\va,t/2)$, although resulting in the desired full mean definition $f^*$ (see \eqref{fbardefs}), would not allow the subsequent evolution equations to depend only on fields at the current time. 

The definitions in \eqref{fbarpdefs} ensure that the full Lagrangian means defined in \eqref{fbardefs} are recovered by setting $t = t^* + T$ in \eqref{fbarpdefs}.

We also define partial mean equivalents of \eqref{Xifullmaps1}-\eqref{Xifullmaps3} (where as before, $\left(\vXip{i\mapsto j}\right)^{-1} = \vXip{j\mapsto i}$)

\begin{align}
\vXip{1\mapsto 2}(\vp(\va,t),t) &= \vpbarp(\va,t)\,, \label{Xipartialmaps1} \\
\vXip{1\mapsto 3}(\vp(\va,t),t) &= \vpstar(\va,t)\,,\label{Xipartialmaps2}\\
\vXip{2\mapsto 3}(\vpbarp(\va,t),t) &= \vpstar(\va,t)\,.\label{Xipartialmaps3}
\end{align}
We now have all of the notation necessary to derive three separate strategies for finding $\barL{f}$ and/or $f^*$, which may be sufficient (see \S \ref{sec:wavemeandecomp}). We briefly summarise each strategy here. In table \ref{tab:strat}, we summarise the dependent fields of each PDE to be solved for the three strategies.\\

\begin{enumerate}
     \item {\bf Strategy 1:} Solve for $\tilde f (\vx,t^*)$ and the map $\vXia{1\mapsto 2}(\vx,t^*)$, then find $\barL{f}(\vx,t^*) = \tilde f((\vXia{1\mapsto 2})^{-1}(\vx,t^*),t^*)$ by a final interpolation step. The variable $\ f^*(\vx,t^*) = \tilde f((\vXia{1\mapsto 3})^{-1}(\vx,t^*),t^*)$ can also be found by solving for $\vXia{1\mapsto 3}(\vx,t^*)$.\\
    \item {\bf Strategy 2:} Solve directly for $\barL{f} (\vx,t^*)$, which requires also solving for the map $\vXia{2\mapsto 1}(\vx,t^*)$. The variable $f^*(\vx,t^*) = \barL{f} ((\vXia{2\mapsto 3})^{-1}(\vx,t^*),t^*)$ can also be found by solving for $\vXia{2\mapsto 3}(\vx,t^*)$.\\
    \item {\bf Strategy 3:} Solve directly for $ f^* (\vx,t^*)$, which requires also solving for the map $\vXia{3\mapsto 1}(\vx,t^*)$. The variable $\barL{f}(\vx,t^*) = f^* ((\vXia{3\mapsto 2})^{-1}(\vx,t^*),t^*)$ can be found by solving for $\vXia{3\mapsto 2}(\vx,t^*)$.\\
\end{enumerate}

\begin{table}
  \begin{center}
\def~{\hphantom{0}}
  \begin{tabular}{lccc}
       \bf{PDE to solve} & \bf{Strategy 1}\hspace{0.5cm}   &   \bf{Strategy 2}\hspace{0.5cm} & \bf{Strategy 3} \\[5pt]
       Scalar equation  & $\tilde{f}_p$ & $\barL{f}_p$ & $f^*_p$\\[3pt]
       Auxiliary map for scalar equation   & --- & $\vXip{2\mapsto 1}$ & $\vXip{3\mapsto 1}$\\[3pt]
       Extra map for interpolation to $\barL{f}$ & $\vXip{1\mapsto 2}$ & --- & $\vXip{3\mapsto 2}$\\[3pt]
       Extra map for interpolation to $f^*$   & $\vXip{1\mapsto 3}$ & $\vXip{2\mapsto 3}$ & ---\\
       
  \end{tabular}
  \caption{Fields to be solved for in each of the three strategies presented.}
  \label{tab:strat}
  \end{center}
\end{table}
\subsection{Strategy 1: Solve at partial trajectory endpoint $\vp(\va,t)$}
The first strategy consists of solving for $\tilde{f}(\vx,t^*)$, the Lagrangian mean along the trajectory of a particle whose position $\vx = \vp(\va,t^*+T)$ is at the trajectory endpoint (as defined in \eqref{fbardefs}).

Taking the time derivative of \eqref{fbarpdefs} at fixed $\va$, and using the dummy variable $\vx = \vp(\va,t)$, we find
\begin{equation}
    \frac{\partial \tilde f_p}{\partial t}(\vx,t) + \vu(\vx,t)\bcdot \nabla \tilde f_p(\vx,t) = f(\vx,t)G(t^* - t)\,.\label{partial_f_evol1}
\end{equation}
where we have used $\vu(\vp(\va,t),t) = \frac{\partial \vp}{\partial t}(\va,t)$ by definition of the flow map. 

Equation \eqref{partial_f_evol1}, along with this initial condition $\tilde f_p(\vx,0) = 0$ (see \eqref{fbarpdefs}) can be solved alongside the governing equations to find $\tilde{f}(\vx,t^*) = \tilde{f}_p(\vx,t^*+T)$. However, we then need to map to the $\vpbar$ coordinates to find $\barL{f}$. We therefore differentiate the definition of $\vXip{1\mapsto 2}$ in \eqref{Xipartialmaps1}, and use the definition \eqref{vpbarpdef} of $\vpbarp(\va,t)$ (setting $\vx = \vp(\va,t)$ as before) to give
\begin{equation}\label{xi12eq}
    \frac{\partial \vxip{1\mapsto 2}}{\partial t}(\vx,t) + \vu(\vx,t)\bcdot \nabla \vxip{1\mapsto 2} (\vx,t) = - \vu(\vx,t)\int_{t^*-T}^t G(t^*-s)\,\d s\,,
\end{equation}
where the perturbation map $\vxip{1\mapsto 2}$ is defined by
\begin{equation}\label{xiXi1}
    \vxip{1\mapsto 2}(\vx,t) =  \vXip{1\mapsto 2}(\vx,t) - \vx\,.
\end{equation}

Initial conditions for \eqref{xi12eq} are given by $\vxip{1\mapsto 2}(\vx,0) = 0$. Evolving \eqref{xi12eq} alongside \eqref{partial_f_evol1} and the governing equations, we can then use $\vXia{1\mapsto 2}(\vx,t^*) = \vXip{1\mapsto 2}(\vx,t^*+T) = \vxip{1\mapsto 2}(\vx,t^*+T) + \vx$ to remap $\tilde f$ to $\barL{f}$, so  that, using \eqref{fbardefs} and writing in terms of a dummy variable $\vx$: 
\begin{equation}\label{strat1interp}
    \barL{f}(\vx,t^*) = \tilde f((\vXia{1\mapsto 2})^{-1}(\vx,t^*),t^*) \,.
\end{equation}
If we would also like to know the mean field $f^*$ defined at the flow trajectory midpoint, we can  solve for $\vXia{1\mapsto 3}(\vx,t^*)$ by differentiating the definition of $\vXip{1\mapsto 3}(\vx,t)$ in \eqref{Xipartialmaps2} and using \eqref{vpstar} to find
\begin{equation}\label{xi13eq}
    \frac{\partial \vxip{1\mapsto 3}}{\partial t}(\vx,t) + \vu(\vx,t) \bcdot \nabla \vxip{1\mapsto 3}(\vx,t) = -\vu(\vx,t)H(t-t^*)\,,
\end{equation}
where $H(\cdot)$ is the Heaviside step function, and the perturbation map $\vxip{1\mapsto 3}$ is defined by 
\begin{equation}
 \vxip{1\mapsto 3}(\vx,t) = \vXip{1\mapsto 3}(\vx,t) - \vx \,.  
\end{equation}
After \eqref{xi13eq} has been evolved to time $t^*+T$, $f^*(\vx,t^*)$ can then be found using $\vXia{1\mapsto 3}(\vx,t^*) = \vXip{1\mapsto 3}(\vx,t^*+T) = \vxip{1\mapsto 3}(\vx,t^*+T) + \vx$ by interpolation from
\begin{equation}\label{strat13interp}
     f^*(\vx,t^*) = \tilde f((\vXia{1\mapsto 3})^{-1}(\vx,t^*),t^*) \,.
\end{equation}
Strategy 1 is the simplest and cheapest strategy, since the evolution equations \eqref{partial_f_evol1}, \eqref{xi12eq} and \eqref{xi13eq} do not involve interpolation at each time step (as will be required by strategies 2 and 3). However, in a complex flow with time-scales similar to or smaller than the averaging interval, the maps $\vXia{12}$ and $\vXia{13}$ can be far from the identity map, making the final interpolation step inaccurate. This will be discussed further in \S \ref{sec:interpolation}, and a case where this interpolation is too complex will be shown in figure \ref{fig:strat13_comparison}. For this reason, KV23 developed a strategy 2, which avoids the need for this interpolation step.

\subsection{Strategy 2: Solve at trajectory partial mean $\vpbarp(\va,t)$}
Following KV23, in strategy 2 we solve directly for $\barL{f}(\vx,t^*)$, the Lagrangian mean  along the trajectory of a particle whose position $\vx = \vpbar(\va,t^*)$ is at the trajectory mean position (as defined in \eqref{fbardefs}).

Taking the derivative with respect to $t$ of \eqref{vpbarpdef}, we define

\begin{equation}\label{ubardef}
    \bar{\vu}_p(\vpbarp(\va,t),t) \equiv \frac{\partial \vpbarp}{\partial t} = \vu(\vp(\va,t),t)\left(1 - \int_{t^*-T}^t G(t^*-s) \,\d s\right)\,.
\end{equation}

We note that $\bar{\vu}_p$ is not related to the Lagrangian mean velocity $\bar{\bar{\vu}}$ defined in \S\ref{sec:Lagrangianvel}. Here, $\bar{\vu}_p$ is found by taking the time derivative of $\vpbarp$ with respect to $t$, whereas the Lagrangian mean velocity is the derivative of $\vpbar$ with respect to $t^*$. 

Then, differentiating the definition of $\vXip{2\mapsto 1}$ in \eqref{Xipartialmaps1} and letting $\vx = \vpbarp(\va,t)$ gives
\begin{equation}
    \frac{\partial \vxip{2\mapsto 1}}{\partial t}(\vx,t) + \bar{\vu}_p(\vx,t)\bcdot \nabla \vxip{2\mapsto 1} (\vx,t) = \vu(\vx + \vxip{2\mapsto 1}(\vx,t),t)\int_{t^*-T}^t G(t^*-s) \,\d s \,, \label{xi21eq}
\end{equation}
where the perturbation map $\vxip{2\mapsto 1}$ is defined by
\begin{equation}
    \vxip{2\mapsto 1}(\vx,t) = \vXip{2\mapsto 1}(\vx,t) - \vx\,,
\end{equation}
with initial condition $\vxip{2\mapsto 1}(\vx,0) = 0$, and from \eqref{ubardef},

\begin{equation}\label{ubardef3}
\bar{\vu}_p(\vx,t) = \vu(\vx + \vxip{2\mapsto1}(\vx,t),t)\left(1 - \int_{t^*-T}^t G(t^*-s) \,\d s\right)\,.
\end{equation}

The evolution of the partial Lagrangian mean scalar $\barL{f}_p$ can be found by taking the time derivative of \eqref{fbarpdefs}, giving
\begin{equation}\label{partial_f_evol2}
    \frac{\partial \barL{f}_p}{\partial t}(\vx,t) + \bar{\vu}_p(\vx,t)\bcdot \nabla \barL{f}_p(\vx,t) = f(\vx + \vxip{2\mapsto 1}(\vx,t),t)G(t^* - t)\,,
\end{equation}
with initial condition $\barL{f}_p(\vx,0) = 0$.

Solving the system of equations \eqref{xi21eq} - \eqref{partial_f_evol2} then directly gives the Lagrangian mean $\barL{f}$. If desired, we can also find $f^*(\vx,t^*)$ by solving for the map $\vXia{2\mapsto 3}(\vx,t^*)$. Differentiating the definition of the map $\vXip{2\mapsto 3}$ in \eqref{Xipartialmaps3}, and setting $\vx = \vpbarp(\va,t)$ leads to
\begin{equation}\label{xi23eq}
    \frac{\partial \vxip{2\mapsto 3}}{\partial t}(\vx,t) + \bar{\vu}_p(\vx,t)\bcdot \nabla \vxip{2\mapsto 3} (\vx,t) = \vu(\vx + \vxip{2\mapsto 1}(\vx,t),t)\left(\int_{t^*-T}^t G(t^*-s) \,\d s - H(t - t^*)\right)\,,
\end{equation}
where the perturbation map $\vxip{2\mapsto 3}$ is defined by
\begin{equation}
    \vxip{2\mapsto 3}(\vx,t) = \vXip{2\mapsto 3}(\vx,t) - \vx\,.
\end{equation}
Then, $f^*(\vx,t^*)$ can be found by interpolation according to
\begin{equation}\label{strat2interp}
     f^*(\vx,t^*) = \barL{f}((\vXia{2\mapsto 3})^{-1}(\vx,t^*),t^*) \,.
\end{equation}
Strategy 2 is intended to eliminate the problems with the final interpolation step in strategy 1 by solving directly at the partial Lagrangian mean position. However, this comes at the expense of a more complicated $RHS$ in the evolution equations \eqref{partial_f_evol2}, \eqref{xi21eq}, and \eqref{xi23eq}, which require interpolation at each time step.

A key disadvantage of strategy 2 pertains to the boundaries of the fluid domain. Since Lagrangian variables are referenced to the trajectory mean position, the equations are posed on a moving domain that will not in general coincide with the fluid domain, making boundary conditions non-trivial -- this is discussed further in \S \ref{sec:BCs}. We therefore now derive a third strategy that enables Lagrangian filtering in more complex and realistic domains. 

\subsection{Strategy 3: Solve at trajectory midpoint $\vpstar(\va,t)$}\label{sec:strat3}
We solve directly for $f^*(\vx,t^*)$, the Lagrangian mean along the trajectory of a particle whose position $\vx = \vp(\va,t^*)$ is at the trajectory midpoint (as defined in \eqref{fbardefs}). For this, we need to solve for the map $\vXia{3\mapsto 1}$, and also for the map $\vXia{3\mapsto 2}$ if we also want to find $\barL{f}(\vx,t^*)$. The derivation is similar to that for strategies 1 and 2, although the first and second halves of the interval must be considered separately. A full derivation is given in Appendix \ref{app:strat3}.

Strategy 3 consists of solving (from \eqref{partial_f_evol31} and \eqref{partial_f_evol32})
\begin{equation}
\frac{\partial f^*_p}{\partial t}(\vx,t) = G(t^* - t)f(\vx + \vxip{3\mapsto 1}(\vx,t),t) - H(t^*-t)\vu(\vx,t)\bcdot \nabla f^*_p(\vx,t)\,,\label{partial_f_evol3}
\end{equation}
with initial conditions $f^*_p(\vx,0) = 0$, along with (from \eqref{xi31zero} and \eqref{xi31eq2})
\begin{equation}\label{xi31eq}
 \frac{\partial \vxip{3\mapsto 1}}{\partial t}(\vx,t) = H(t-t^*)\vu(\vx + \vxip{3\mapsto 1}(\vx,t),t)\,,
\end{equation}
with initial conditions $\vxip{3\mapsto 1}(\vx,0) = 0$. If $\barL{f}(\vx,t^*)$ is required, then we also solve (from \eqref{xi32eq1} and \eqref{xi32eq2})
\begin{equation}
\begin{split}\label{xi32eq}
\frac{\partial \vxip{3\mapsto 2}}{\partial t}(\vx,t) &= \vu(\vx + \vxip{3\mapsto1}(\vx,t),t)\left(H(t-t^*) - \int_{t^*-T}^t G(t^*-s)\,\d s\right) \\&- H(t^*-t)\vu(\vx,t)\bcdot \nabla \vxip{3\mapsto 2} (\vx,t)\,,
\end{split}
\end{equation}
with initial conditions $\vxip{3\mapsto 2}(\vx,0) = 0$. Then, $\barL{f}(\vx,t^*)$ can be found using $\vXia{3\mapsto 2}(\vx,t^*) = \vXip{3\mapsto 2}(\vx,t^*+T) = \vxip{3\mapsto 2}(\vx,t^*+T) + \vx$ from
\begin{equation}\label{strat3interp}
     \barL{f}(\vx,t^*) = f^*((\vXia{3\mapsto 2})^{-1}(\vx,t^*),t^*) \,.
\end{equation}
Like strategy 1, strategy 3 requires a final interpolation step if $\barL{f}$ is required (rather than $f^*$). However, this final interpolation, performed using $\vXia{3\mapsto 2}$, is likely to be much more accurate than that in strategy 1, since the trajectory  mean and midpoint positions differ only by the wave perturbation (see figure \ref{fig:trajectorydiagram}). This will be demonstrated in figure \ref{fig:strat13_comparison}. 

\subsection{Boundary conditions}\label{sec:BCs}
In this study, we consider the simplest case of a doubly periodic domain. This is simple to implement as the Lagrangian mean equations for each strategy are constructed so that periodic state fields of the simulation lead to periodic Lagrangian mean fields (the normalisation condition \eqref{normalisation} is essential for this to be the case). However, some of the equations can also be solved in more complex domains. 

Any fluid in a domain with open (non-periodic) boundaries will contain trajectories that exit the domain, so all definitions of Lagrangian means for these trajectories will be undefined and Lagrangian mean fields cannot be calculated over the full domain. However, the equations of strategies 1 and 3 (\eqref{partial_f_evol1}, \eqref{xi12eq}, \eqref{xi13eq}, and \eqref{partial_f_evol3} - \eqref{xi32eq}) can be straightforwardly solved in any domain with fixed boundaries. Each of the equations for Lagrangian fields in these strategies contains an advective derivative term $\vu \bcdot \nabla$, indicating that a boundary condition is needed. However, in a fixed bounded domain with normal $\bm{n}$, the velocity satisfies $\vu \bcdot \bm{n} = 0$, so the normal part of the advective term vanishes at the boundary and no boundary conditions on the Lagrangian fields are necessary. After having solved for $\tilde{f}$ (strategy 1) or $f^*$ (strategy 3), the final interpolation can then be carried out to find $\barL{f}$, although there is no guarantee that $\barL{f}$ can be defined at every point in the domain (i.e. for a domain $\mathcal{D}$ and some $\mathbf{y} \in \mathcal{D}$, there may not exist $\vx \in \mathcal{D}$ such that $\vXia{3\mapsto 2}(\vx,t) = \mathbf{y}$).

In contrast, strategy 2 cannot easily be used in fixed bounded domains. Since the Lagrangian fields are defined on the image of the partial mean flow map $\vpbarp$, the PDEs are posed on a domain with moving boundaries, leading in general to a free boundary problem. There is no guarantee that the Lagrangian mean position itself lies in the fluid domain (unless it is convex), or that a given location in the fluid domain is the Lagrangian mean position of some trajectory, so $\barL{f}$ may not be defined everywhere. 

Strategy 2 can however be straightforwardly used in domains that have at most one non-periodic dimension, along which the boundaries must align with a constant coordinate surface in that dimension (i.e. one set of straight and parallel boundaries in Euclidean space). In this case, the image of the mean flow map coincides with the fluid domain. Boundary conditions are not needed, since trajectories stay on the boundary and thus $\vu(\vx,t) \bcdot \bm{n} = 0 \Rightarrow \vu(\vx + \vxip{2\mapsto1},t)\bcdot \bm{n} =0  \Rightarrow \vu_p(\vx,t) \bcdot \bm{n} =0$.

\section{Numerical model}\label{sec:numericalmodel}

We now demonstrate our filtering approach using a single layer rotating shallow water system, which permits both geostrophic turbulence and Poincar\'e waves. 

Similarly to KV23, we use the rotating shallow water equations in a doubly periodic domain. However, in order to have more flexibility over the chosen wavenumbers of Poincar\'e waves, we use the modified shallow water (MSW) equations introduced by \citet{buhlerShallowWaterModelThat1998}. These equations were developed for the very purpose of providing a simple test-bed for wave--mean decompositions, without the added complication of steepening Poincar\'e waves that occurs in the regular shallow water equations \citep{buhlerShallowWaterModelThat1998}. The MSW equations behave similarly to the shallow water equations, and the equations are identical when linearised about a state of rest. In our case, we want a flow that contains a slowly varying `mean' component alongside a wave field, and are agnostic to the physicality of the flow. We work with non-dimensional quantities -- see KV23 for details of the non-dimensionalisation. The flow equations are
\begin{align}
    \frac{\partial \vu}{\partial t} + \vu \bcdot \nabla \vu + \frac{1}{Ro}\hat{\bm{z}}\times \vu &= -\frac{1}{Fr^2}\mathcal{F}(h)\nabla h\,, \label{SWmom}\\
    \frac{\partial h}{\partial t} + \nabla \bcdot (\vu h) &= 0 \,,\label{SWheight}
\end{align}
where $\vu = (u,v,0)$ is the velocity, $h(x,y)$ is the height, and motion is on an $x,y$ plane perpendicular to the vertical unit vector $\hat{\mathbf{z}}$. $\mathcal{F}(h) = 1$ for standard shallow water, and 
\begin{equation}\label{MSWF}
    \mathcal{F}(h) = \frac{1}{h^3}
\end{equation}
for MSW. The non-dimensional parameters are the Froude and Rossby numbers $Fr$ and $Ro$, where $Fr$ is the non-dimensional inverse phase speed of linear gravity waves unaffected by rotation, and $Ro$ represents the ratio of inertial to Coriolis forces. Throughout, we take $Fr = 0.3$ and $Ro = 0.4$.

The flow is initialised with the output of an incompressible two-dimensional Navier–Stokes simulation in a fully-developed turbulent state, with height $h$ set to be initially in geostrophic balance (as in KV23)
\begin{equation}\label{geostrophy}
    \frac{1}{Ro}\hat{\bm{z}}\times \vu = -\frac{1}{Fr^2}\nabla h\,,
\end{equation}
and is allowed to evolve freely. The non-dimensionalisation of the height is such that $h = 1 + \eta$, where the height perturbation $\eta = 0$ for a flow at rest. For $\eta \ll 1$, the MSW term in \eqref{MSWF} therefore scales as $\mathcal{F}(h) = 1 + O(\eta)$, and MSW approximates standard shallow water. For a flow in geostrophic balance, $\eta \sim Fr^2/Ro$ (see \eqref{geostrophy}), so $Fr^2/Ro \ll 1$ is the condition for such a geostrophic flow to behave similarly to a standard shallow water flow. Here, $Fr^2/Ro = 0.225$, and we find that this is sufficient to prevent any spurious behaviour from the shallow water modification. 

We also superimpose a linear Poincar\'e wave on this initial condition and allow it to evolve alongside and interact with the geostrophic turbulence. The linearisation of the MSW equations \eqref{SWmom} -\eqref{SWheight} is identical to the original shallow water system. Linear wave solutions have frequency $\omega$ satisfying the dispersion relation
\begin{equation}\label{disprel}
    \omega^2 = \frac{1}{Ro^2} + \frac{|\bm{k}|^2}{Fr^2}\,,
\end{equation}
where $\bm{k} = (k,l)$ is the wavenumber. The height perturbation $\eta$ of the waves scales as $A Ro$, where $A$ is the maximum amplitude of the vorticity of the initialised wave. We take $A = 0.5$, so that $A Ro = 0.2$, and the waves are also sufficiently linear to not be obviously affected by the MSW term \eqref{MSWF}, aside from their lack of nonlinear steepening as intended. The mode-1 ($|\mathbf{k}| = 1$) wave has frequency $\omega = 4.17$.

Starting from this initial condition, we evolve the MSW equations alongside the Lagrangian mean equations using a pseudo-spectral solver as in KV23, with a fourth order Runge Kutta scheme for the advective terms \citep{bakerSWGLMsoftware24}. We use a non-dimensional domain size of $2\pi \times 2\pi$, with 256 gridpoints in the $x$ and $y$ directions. A hyperviscous (Laplacian to the power four) term is added to the momentum equation \eqref{SWmom} to remove energy at small scales. A hyperviscous term can also be added to the Lagrangian mean equations, which is found to be necessary for numerical stability when integrating strategy 1 over long time intervals. However, this is not necessary and is found to introduce error in the resulting Lagrangian mean for strategies 2 and 3, where the forcing terms on the $RHS$ of the scalar equations \eqref{partial_f_evol2} and \eqref{partial_f_evol3} seem to stabilise the simulations, so the Lagrangian mean equations of strategies 2 and 3 are run without any viscous terms. 

We implement each of strategies 1, 2, and 3, but show results from only strategies 1 and 3, since the results of strategies 2 and 3 are visually identical (although their difference is quantified in \S \ref{sec:meanresults}), but strategy 3 is faster (see Appendix \ref{app:expense}). Unless otherwise stated, the equations are run using strategy 3 for an averaging period of $2T = 40$, over which time the mean and wave components of the flow both evolve. The weight function $G$ is truncated to this finite interval, and renormalised to ensure that \eqref{normalisation} holds exactly over the interval. In the case of a low-pass with weight function given by \eqref{lowpass},
\begin{equation}
    \int_{t^*-T}^{t^* + T}G_{\mathrm{LP}}(t^*-s)\,\d s \rightarrow 1 \hspace{0.5cm} \mathrm{as} \hspace{0.5cm}\omega_c T \rightarrow \infty\,,
\end{equation}
so when $\omega_cT$ is sufficiently large, the normalisation requirement still approximately holds. Unless otherwise stated, we use a low-pass cut-off frequency of $\omega_c = 2$, so that $\omega_c T = 40$, and $\int_{t^*-T}^{t^* + T}G_{\mathrm{LP}}(t^*-s)\,\d s = 1.01$. Appendix \ref{app:intervaltimes} shows the impact of changing $T$. 

The scalar field to be averaged is the relative vorticity
\begin{equation}
    \zeta = \frac{\partial v}{\partial x} - \frac{\partial u}{\partial y}\,.
\end{equation}

First, we show results for the Lagrangian mean of the vorticity and compare the different strategies. We then explain the various ways that the flow can be decomposed into wave and mean components, before showing results for these decompositions. 

\section{Results: Lagrangian mean}\label{sec:meanresults}

\begin{figure}
    \centering
    \includegraphics[width=\textwidth]{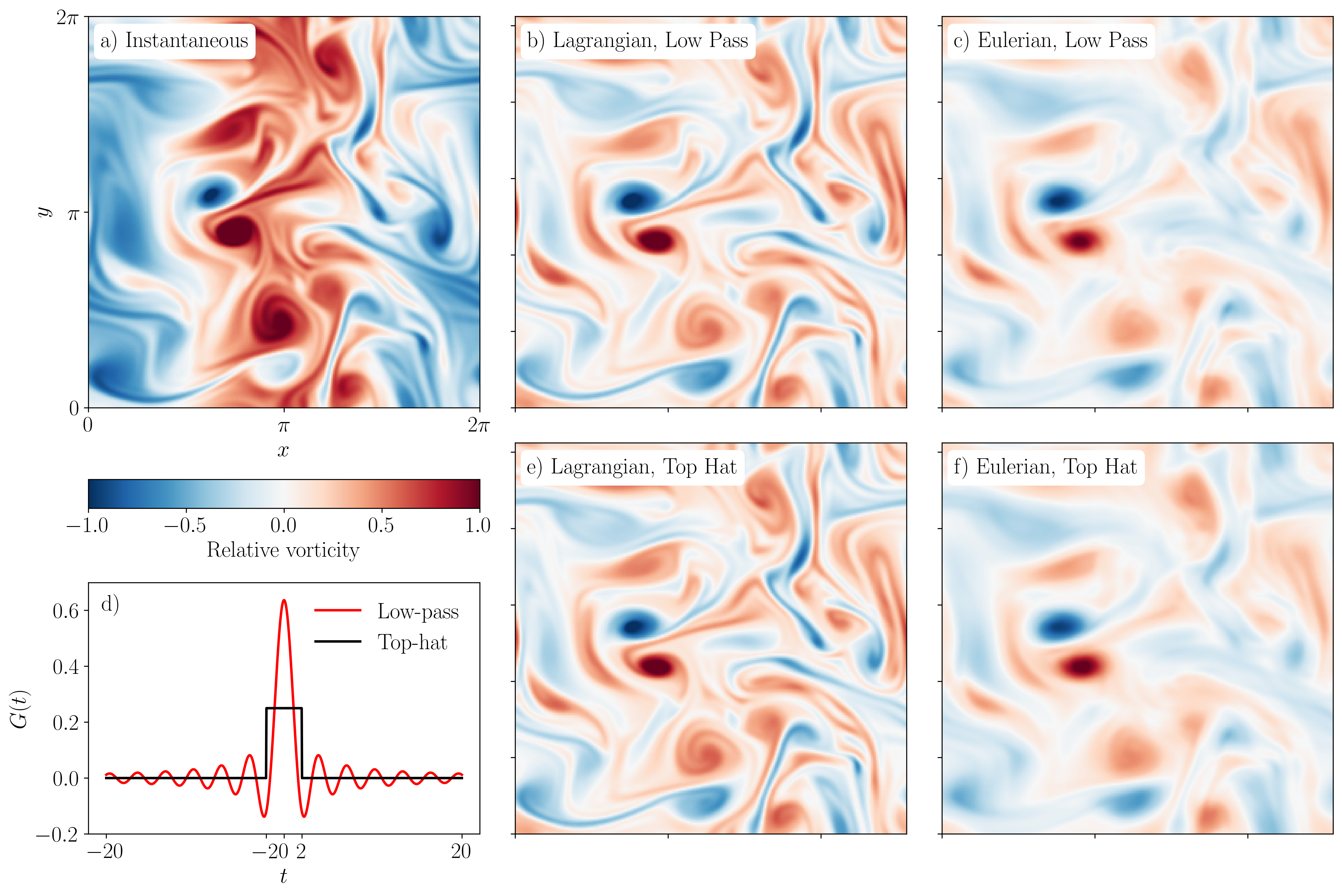}
    \caption{Shallow water relative vorticity for a simulation over 40 time units ($T=20$). The mode-1 wave frequency is $\omega = 4.17$, and the low-pass filters use a cut-off frequency of $\omega_c = 2$. a) Instantaneous vorticity at the interval midpoint $t^* = 20$. b) Lagrangian and c) Eulerian low-pass at $t^* = 20$. e) Lagrangian and f) Eulerian top-hat mean at $t^* = 20$, computed over the interval $[18,22]$, i.e. $T = 2$. d) $G(t)$ for the low-pass and top-hat means, showing that $T = 2$ is an appropriate averaging interval for the top-hat to compare it to the low-pass. The directory including the Jupyter notebook that generated this figure can be accessed at \url{https://cocalc.com/share/public_paths/bdc0d1617e113644a25e3ba4c0b91b8fad20701f/Figure-2}.}
    \label{fig:mean_comparison}
\end{figure}

Figure \ref{fig:mean_comparison}a shows the instantaneous vorticity at the midpoint of the averaging interval for comparison to various means. Whilst there is a high amplitude mode-1 wave present in the instantaneous vorticity, this wave is removed by the averaging procedures in figures \ref{fig:mean_comparison}b, \ref{fig:mean_comparison}c, \ref{fig:mean_comparison}e, and \ref{fig:mean_comparison}f.

Figures \ref{fig:mean_comparison}b and \ref{fig:mean_comparison}c show the Lagrangian and Eulerian low-pass means of vorticity. The Lagrangian low-pass retains more of the intensity of the vortices than the Eulerian low-pass, since the effect of the large amplitude wave displacement on the background turbulence leads to a blurring of the vortices in the Eulerian low-pass. This low-pass is calculated over an interval of 40 time units ($T = 20$), and the corresponding weight function $G(t)$ is shown in figure \ref{fig:mean_comparison}d. The root-mean-squared difference between the Lagrangian mean vorticity in figure \ref{fig:mean_comparison}b when calculated with strategies 2 and 3 is 0.003, with a maximum difference of 0.03. 

Also shown in figure \ref{fig:mean_comparison}d is $G(t)$ for a top-hat mean with a comparable averaging time-scale of $T=2$. Figures \ref{fig:mean_comparison}e and \ref{fig:mean_comparison}f show the corresponding Lagrangian and Eulerian top-hat means at the same value of $t^*$ as for the low-pass. Whilst there are not qualitative differences between the top-hat and low-pass mean vorticity, there are differences that are evident when the fields are viewed as a time series.

\begin{figure}
    \centering
    \includegraphics[width=\textwidth]{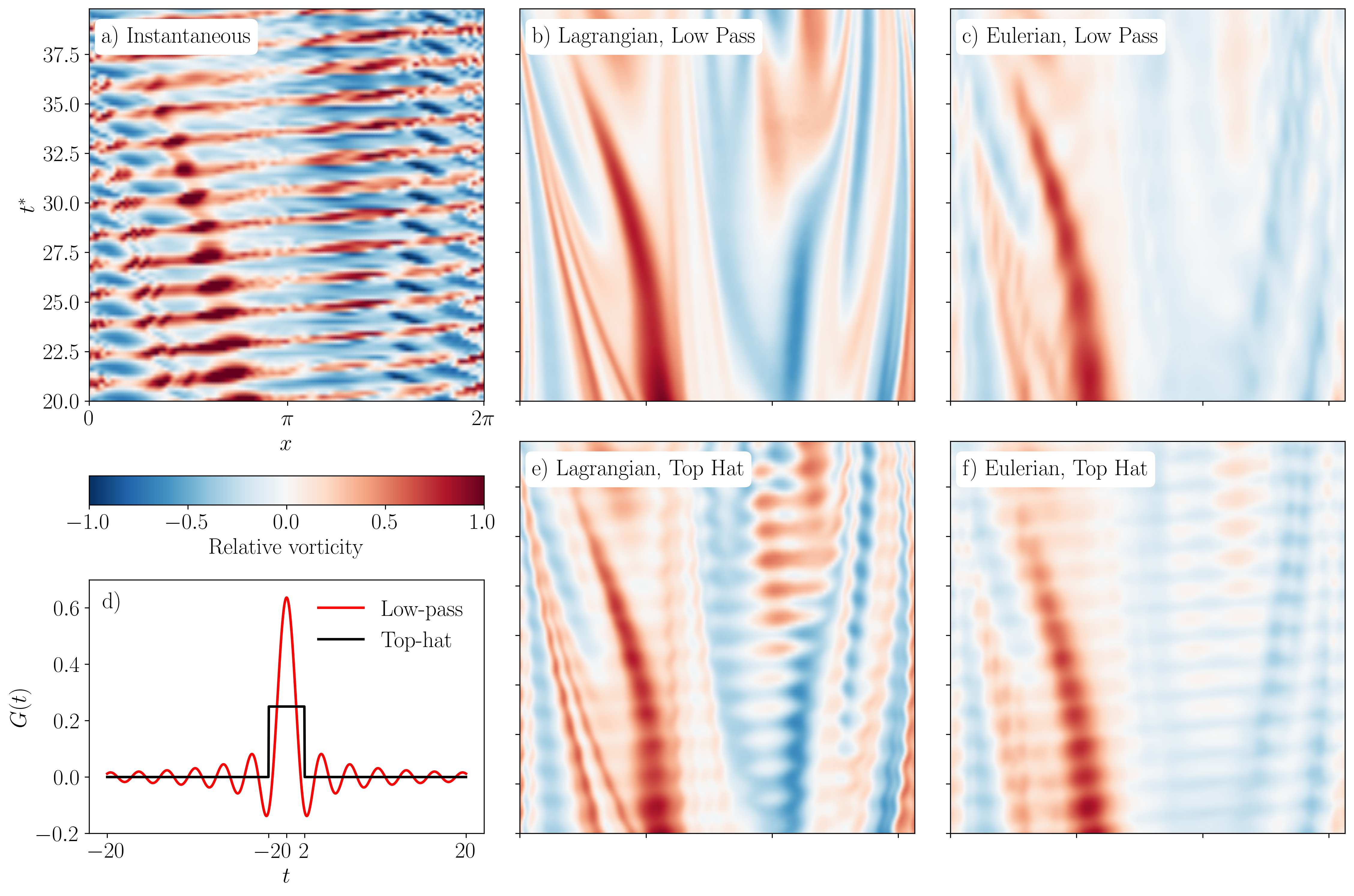}
    \caption{As in figure \ref{fig:mean_comparison}, showing the time ($t^*$) evolution of each field at $y = 2.8$. The directory including the Jupyter notebook that generated this figure can be accessed at \url{https://cocalc.com/share/public_paths/bdc0d1617e113644a25e3ba4c0b91b8fad20701f/Figure-3}.}
    \label{fig:mean_comparison_hovmuller}
\end{figure}

Figure \ref{fig:mean_comparison_hovmuller} shows the same fields as figure \ref{fig:mean_comparison} over a time series in $t^*$. We note here that figure \ref{fig:mean_comparison_hovmuller} is not showing the evolution in $t$ of the partial Lagrangian mean fields described in \S \ref{sec:formulation}. Instead, for each value of $t^*$, a set of Lagrangian mean equations are solved over the interval $[t^*-T,t^*+T]$, and the values of the full Lagrangian means (referenced to time $t^*$) are shown. 

Whilst the oscillations at the wave frequency are removed by the Lagrangian low-pass in figure \ref{fig:mean_comparison_hovmuller}b, they are still evident in the top-hat mean shown in figure \ref{fig:mean_comparison_hovmuller}e. This is because the top-hat mean is less selective in the frequencies that are filtered. In this simple test case, we could have chosen the averaging period of the top-hat to exactly be the period of the wave to better remove this wave signal. However, in the general case of a continuous spectrum of waves, the top-hat mean would not be able to perfectly remove all waves with a given cut-off frequency. 

The Eulerian low-pass mean in figure \ref{fig:mean_comparison_hovmuller}c is more effective in removing the wave oscillations than the Lagrangian top-hat, but, as in figure \ref{fig:mean_comparison}c, the resulting mean flow is blurred. The Eulerian top-hat in figure \ref{fig:mean_comparison_hovmuller} suffers from both blurring and residual wave signal. Hereafter, we focus on the low-pass filter, as it gives more control over the frequencies to remove, and consider the relative merits of strategies 1 and 3. 

\begin{figure}
    \centering
    \includegraphics[width=\textwidth]{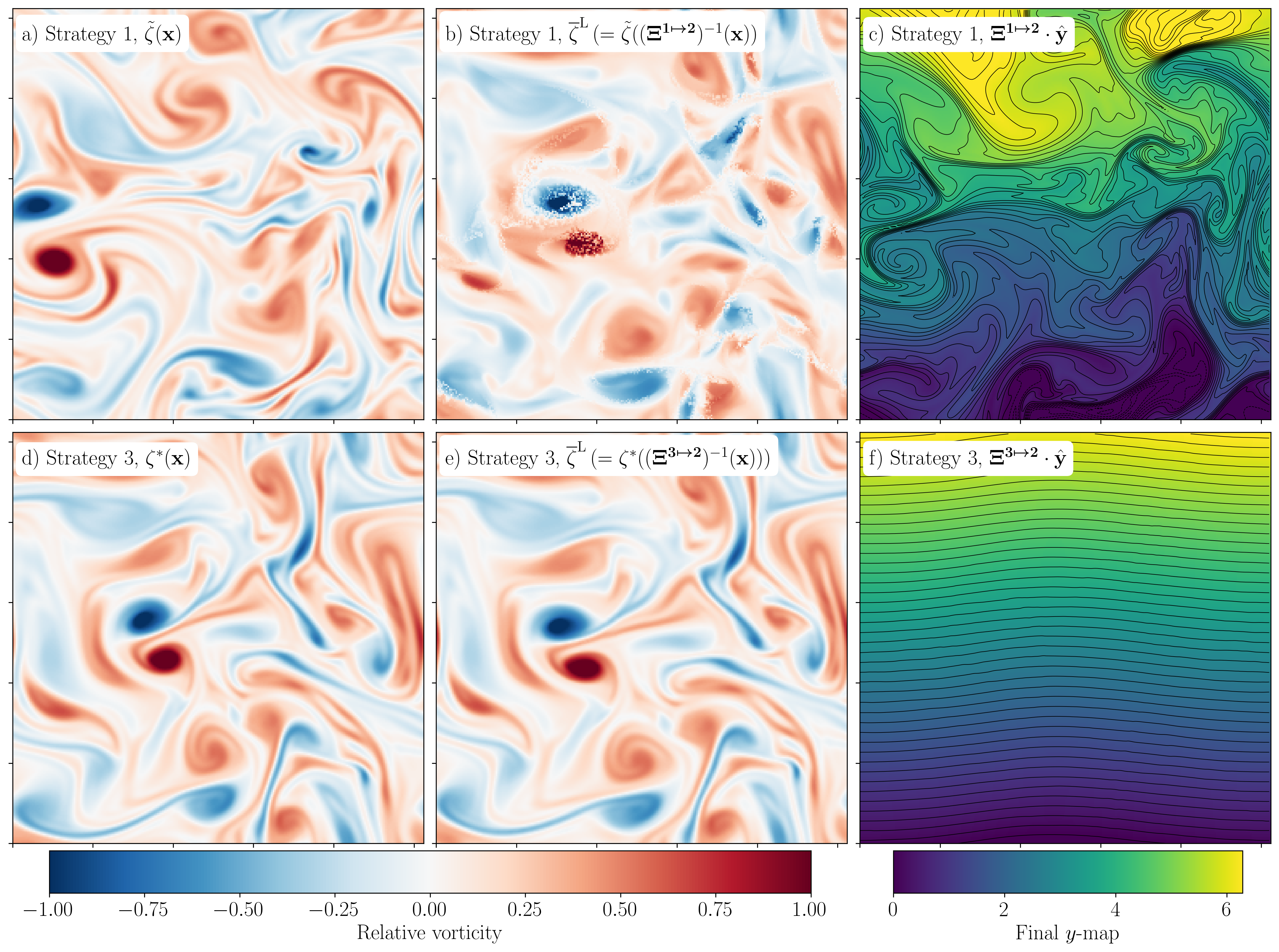}
    \caption{Comparison of calculation of $\barL{\zeta}$ using strategies 1 and 3 with $T=20$. a) $\tilde{\zeta}$, found using strategy 1, b) $\barL{\zeta}$, found by remapping $\tilde{\zeta}$ using $\vXia{1 \mapsto 2}$, and c) the $y$ component of $\vXia{1 \mapsto 2}$. d) $\zeta^*$, found using strategy 3, e) $\barL{\zeta}$, found by remapping $\zeta^*$ using $\vXia{1 \mapsto 3}$, and f) the $y$ component of $\vXia{1 \mapsto 3}$. $x$ and $y$ axes correspond to $x$ and $y$ coordinates of the full domain. The directory including the Jupyter notebook that generated this figure can be accessed at \url{https://cocalc.com/share/public_paths/bdc0d1617e113644a25e3ba4c0b91b8fad20701f/Figure-4}.}
    \label{fig:strat13_comparison}
\end{figure}

Figure \ref{fig:strat13_comparison}a shows the direct output $\tilde{\zeta}$ of strategy 1, and figure \ref{fig:strat13_comparison}d the direct output $\zeta^*$ of strategy 3 (which are rearrangements of each other). To find $\barL{\zeta}$, these fields are remapped to the trajectory mean coordinate, using $\vXia{1\mapsto 2}$ for strategy 1 and $\vXia{3\mapsto 2}$ for strategy 3. The $y$ components of these maps are shown in figures \ref{fig:strat13_comparison}c and \ref{fig:strat13_comparison}f respectively, and the resulting $\barL{\zeta}$ in figures \ref{fig:strat13_comparison}b and \ref{fig:strat13_comparison}e. The map $\vXia{1\mapsto 2}$ is complicated as it represents the motion of the flow between the trajectory mean and end positions, and the resulting $\barL{\zeta}$ is poorly interpolated. However, $\vXia{3\mapsto 2}$ differs from the identity only by the wave perturbation, and therefore results in a clean interpolation to $\barL{\zeta}$. Therefore, strategy 3 is preferred over strategy 1 when the mean flow varies significantly over the averaging interval, although in a flow with a more complex `wave' component, the final mapping of strategy 3 may still be too complicated. However, over a shorter averaging interval, where the interpolating map shown in figure \ref{fig:strat13_comparison}c is simpler, strategy 1 may be more accurate than strategy 3, since strategy 3 can accumulate interpolation errors at each time step. We discuss this further in \S \ref{sec:interpolation}. The complexity of the strategy 1 mapping in figure \ref{fig:strat13_comparison}c is also partly due to the long averaging interval used -- the use of a filter that is more localised in time than the strict low-pass (such as a Butterworth or Gaussian filter) would allow a shorter averaging interval, and correspondingly a less complex final interpolation in strategy 1 (and also in the forcing terms of the strategy 2 and 3 equations). 

Having found the Lagrangian mean of a flow, we now consider how to define the wave-like component of the flow. 

\section{Lagrangian wave formulation}\label{sec:wavemeandecomp}

There are several ways to define the wave-like component of the flow. The first (and perhaps most common) is to define waves as high frequency perturbations at a fixed point. If the Eulerian mean of some scalar $f$ is given by $\overline{f}^\mathrm{E}(\vx,t)$, then the Eulerian wave perturbation is defined as
\begin{equation}\label{Eulerian_wave}
    \overline{f}^\mathrm{w}_\mathrm{E}(\vx,t) = f(\vx,t) - \overline{f}^\mathrm{E}(\vx,t)\,,
\end{equation}
where superscript $\mathrm{w}$ represents the wave component. However, the Lagrangian mean is more effective than the Eulerian mean for recovering a mean flow in the presence of large amplitude waves (e.g. figures \ref{fig:mean_comparison} and \ref{fig:mean_comparison_hovmuller}), or when the waves are significantly Doppler shifted \citep{Shakespeare2021a}. We should therefore define waves to be high frequency motions in the Lagrangian frame, and mean flows to be low frequency motions in the Lagrangian frame, with some appropriate cut-off frequency separating the two. By this definition, mean flows must not necessarily be balanced in the sense of geostrophic balance, or even slowly varying in the Eulerian reference frame. 

However, this definition of a wave is still not precise enough. We have the option to define waves in either a `Semi-Eulerian' or a Lagrangian way. We may say a wave perturbation is:\\

\begin{enumerate}
    \item {\bf Eulerian}: an Eulerian high frequency perturbation at a fixed point.
    \item {\bf Semi-Eulerian}: a Lagrangian high frequency perturbation at a fixed point.
    \item {\bf Lagrangian}: a Lagrangian high frequency perturbation following a particle.\\
\end{enumerate}

The difference between these viewpoints stems from the fact that waves impact the flow in two ways: through changing the value of a scalar as seen by a flow-following particle, and through displacing the mean flow. The semi-Eulerian definition of the wave-field encompasses these two effects, whereas the Lagrangian wave-field only represents the changes in value of a scalar on a particle due to the wave. 

Before posing these decompositions mathematically, we consider a simple example to elucidate the difference between them. Consider a 2D $(x,z)$ background flow with uniform velocity $\mathbf{U} = (U_0,0)$ and buoyancy $B(z)$ that is stably stratified. Steady internal lee waves can propagate on this base state, giving total buoyancy $b$ and horizontal $x$-velocity $u(x,z)$ of the form
\begin{align}
    b(x,z) &= B(z) + b'(x,z)\,, \label{beq}\\
    u(x,z) &= U_0 + u'(x,z)\,. \label{ueq}
\end{align}
Lee waves are generated by flow over topography in the ocean and atmosphere, and are phase-locked to topography such that they are steady in the rest frame, hence $u$ and $b$ are independent of time. The variables $b(x,z)$ and $u(x,z)$ are therefore unchanged by an Eulerian mean, and the Eulerian buoyancy and velocity wave perturbations are zero.

The Lagrangian means of $b(x,z)$ and $u(x,z)$, when filtered with an appropriate cut-off frequency that is lower than the intrinsic wave frequency, are $B(z)$ and $U_0$ respectively, thus the semi-Eulerian wave perturbations are $b'$ and $u'$.

In the absence of diffusion, buoyancy is a conservative tracer satisfying
\begin{equation}
    \frac{D b}{Dt} = \frac{\partial b}{\partial t} + \vu \bcdot \nabla b = 0\,,
\end{equation}
thus buoyancy is constant following a particle, and the Lagrangian buoyancy wave perturbation is zero. The wave velocity $u$ is not constant following a particle, therefore the Lagrangian velocity perturbation is non-zero (and unknown for now). 

\subsection{Semi-Eulerian wave definition}
The semi-Eulerian wave-field is defined to be the instantaneous field minus the Lagrangian mean field at a fixed spatial location. The mean field is given by the Lagrangian weighted mean $\barL{f}$, which, as discussed in \S\ref{sec:definingmean}, contains no wave signal when weighted with the appropriate frequency filter. We define
\begin{equation}\label{Wave_E_def}
    f_\mathrm{S-E}^\mathrm{w}(\vx,t^*) = f(\vx,t^*) - \barL{f}(\vx,t^*)\,,
\end{equation}
where the subscript $\mathrm{S-E}$ denotes a semi-Eulerian wave definition. 

\subsection{Lagrangian wave definitions}
The Lagrangian wave-field is defined as the Lagrangian high frequency perturbation on a trajectory. We have two further options for how this is itself defined -- either at the Lagrangian trajectory midpoint, or at the Lagrangian mean position:
\begin{align}
    f_\mathrm{L1}^\mathrm{w}(\vp(\va,t^*),t^*) &= f(\vp(\va,t^*),t^*) - f^*(\vp(\va,t^*),t^*)\\
    &= f(\vp(\va,t^*),t^*) - \barL{f}(\vpbar(\va,t^*),t^*)\, , \label{L1_w}\\
    f_\mathrm{L2}^\mathrm{w}(\vpbar(\va,t^*),t^*) &=f(\vp(\va,t^*),t^*) - \barL{f}(\vpbar(\va,t^*),t^*)\,.\label{L2_w}
\end{align}
Using the map $\vXia{3\mapsto 2}$ defined by $\vXia{3\mapsto 2}(\vp(\va,t^*),t^*) = \vpbar(\va,t^*)$ (c.f. \eqref{Xifullmaps1}-\eqref{Xifullmaps3} ), and letting a dummy variable $\vx = \vp(\va,t^*)$ in \eqref{L1_w} and $\vx = \vpbar(\va,t^*)$ in \eqref{L2_w} we obtain the alternative forms
\begin{align}
    f_\mathrm{L1}^\mathrm{w}(\vx,t^*) &= f(\vx,t^*) - f^*(\vx,t^*)\\
    &= f(\vx,t^*) - \barL{f}(\vXia{3\mapsto 2}(\vx,t^*),t^*)\, , \label{L1_w_x}\\
    f_\mathrm{L2}^\mathrm{w}(\vx,t^*) &= f((\vXia{3\mapsto 2})^{-1}(\vx,t^*),t^*) - \barL{f}(\vx,t^*) \label{L2_w_x}
\end{align}
Note that the two definitions \eqref{L1_w_x}-\eqref{L2_w_x} are just rearrangements of each other such that $f_\mathrm{L1}^\mathrm{w}(\vx,t^*) = f_\mathrm{L2}^\mathrm{w}(\vXia{3\mapsto 2}(\vx,t^*),t^*)$. 

\subsection{Comparing wave definitions}
Having defined four different wave-fields (one Eulerian \eqref{Eulerian_wave}, one semi-Eulerian \eqref{Wave_E_def} and two Lagrangian \eqref{L1_w_x}-\eqref{L2_w_x}), we now consider the features of each, with a focus on the semi-Eulerian/Lagrangian definitions, having already motivated the Lagrangian over the Eulerian mean. We note that the wave definitions given here do not depend on the strategy with which they are calculated. 

The semi-Eulerian definition \eqref{Wave_E_def} is perhaps the most straightforward to understand, since the wave is defined as `what is left when you remove the Lagrangian mean field'. When it is desirable to write the total field as a sum of mean and wave components at the same spatial location, as is done in the lee wave example \eqref{beq}-\eqref{ueq}, this is the most helpful decomposition. However, although the two terms on the $RHS$ of \eqref{Wave_E_def} are defined at the same spatial location, the instantaneous field $f$ is evaluated at the position $\vx$ which is not necessarily on the path of the particle whose mean position is $\vx$, and whose mean is evaluated in the second term. Hence, we are subtracting the mean of particle from the instantaneous value of a different particle. As in \S \ref{sec:definingmean}, the Lagrangian low-pass filter applied to the wave field would ideally return zero, and this is not the case for the semi-Eulerian wave-field when filtered along trajectories of either the original flow $\vp$ or the mean flow $\vpbar$. 

However, the Lagrangian definitions do have this property, and it can be shown that (assuming a simple band-pass filter as described in \S\ref{sec:definingmean}) the first Lagrangian wave-field is zero when low-pass filtered along the original flow paths, and the second is zero when low-pass filtered along the mean flow paths. For comparison with the well-known notation of \citet{andrewsExactTheoryNonlinear1978}, we note that the second Lagrangian wave definition \eqref{L2_w_x} corresponds to their Lagrangian disturbance quantities with a superscript $l$ (e.g. their equation 2.11), although their Eulerian average (such that $\overline{\,f^l\,}^{\,\mathrm{E}} = 0$) is replaced in our case with a Lagrangian average along mean flow trajectories since we do not assume separation of time-scales (see Appendix \ref{app:timescales}). 

In the first Lagrangian wave definition \eqref{L1_w_x}, a deformation of the mean field that includes the impact of the wave disturbance of the mean field ($f^*$) is subtracted from the instantaneous field to give a wave component $f_\mathrm{L1}^\mathrm{w}$ that documents only the changes to the value of the field seen by a particle. This is the wave field that is found by filtering methods that use particle tracking with particles seeded at the reference time $t^*$, such as \citet{Shakespeare2021a}, which directly find $f^*$ as the mean field (although such methods usually track with horizontal velocities only, so only approximate $f^*$). We will later see that this wave decomposition can give a much clearer view of the wave field than the semi-Eulerian definition, since the wave component does not include the wave-displaced mean flow. However $f^*(\vx,t)$ is not the true mean field, as it includes a wave signal (see figure \ref{fig:hovmoller}c later).

The second Lagrangian wave definition \eqref{L2_w_x} is similar to the first, but to find the wave field the instantaneous total field must first be deformed to remove the effect of wave displacement, before the mean field is subtracted. Therefore, neither of the Lagrangian descriptions give a decomposition that can be written as \textit{wave + mean = total} at a fixed spatial location.

\section{Results: Lagrangian waves}\label{sec:waveresults}

\begin{figure}
    \centering
    \includegraphics[width=\textwidth]{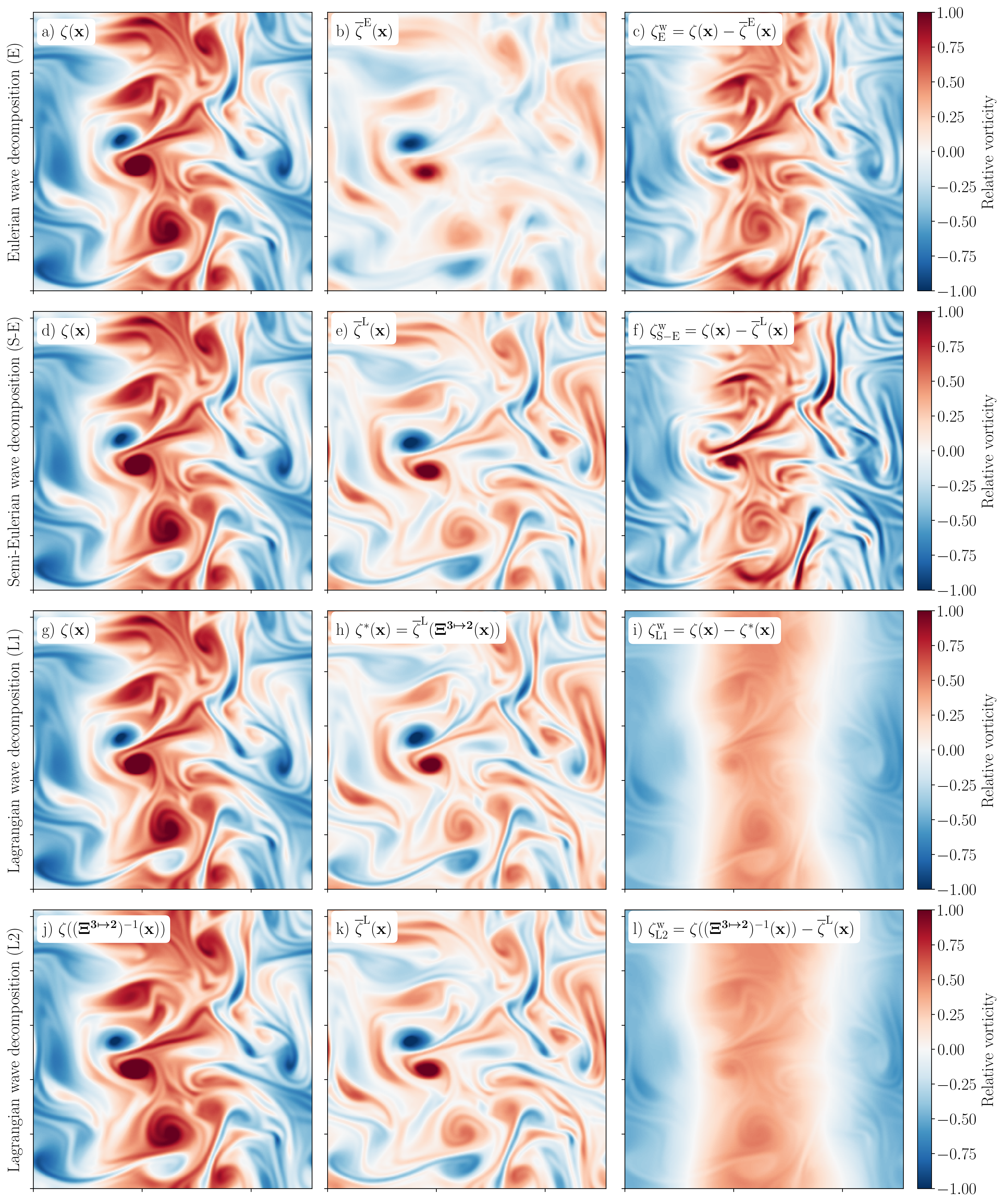}
    \caption{The four different wave decompositions: (top) Eulerian, (second row) semi-Eulerian, (third row) Lagrangian first definition, and (bottom) Lagrangian second definition. For each row, the middle `mean' field is subtracted from the left `instantaneous' field to give the right `wave' field. The flow parameters are as for figure \ref{fig:mean_comparison}, and strategy 3 is used. The directory including the Jupyter notebook that generated this figure can be accessed at \url{https://cocalc.com/share/public_paths/bdc0d1617e113644a25e3ba4c0b91b8fad20701f/Figure-5}.}
    \label{fig:wave_decomp}
\end{figure}

Figure \ref{fig:wave_decomp} shows the four different wave decompositions discussed above for the same example as in figure \ref{fig:mean_comparison}, where in each case the left column minus the middle column gives the wave perturbation in the right column. Both the Eulerian and semi-Lagrangian wave definitions give a wave that has a significant signature of the turbulent mean flow. In the Eulerian case, this is because the mean flow is blurred by the high amplitude wave perturbations, and in the semi-Eulerian case this is because the deformation of the mean field by the high amplitude wave is included in the wave definition. 

\begin{figure}
    \centering
    \includegraphics[width=\textwidth]{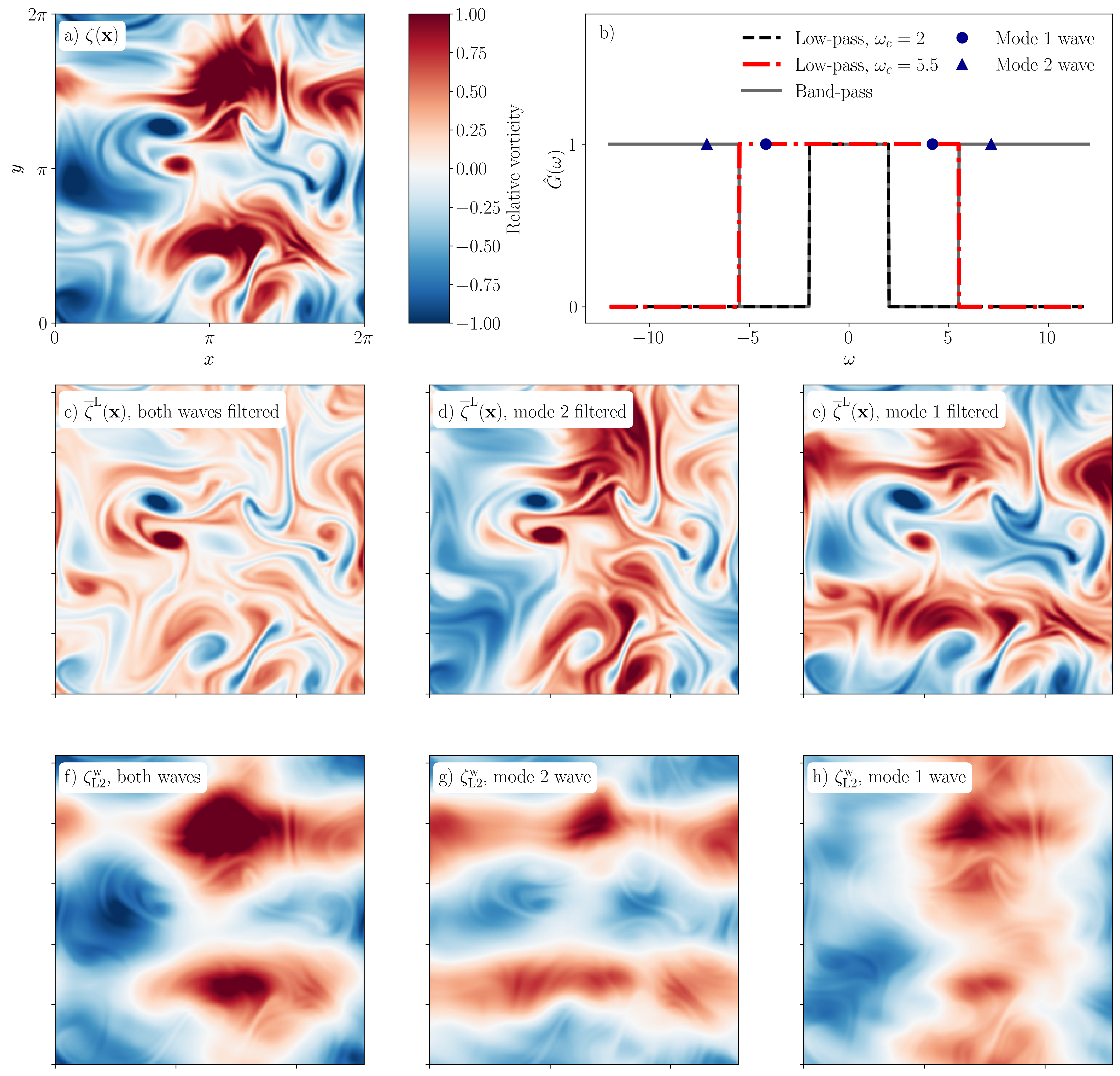}
    \caption{An example of different frequency filters with corresponding functions $\hat{G}(\omega)$ shown in panel b. a) Instantaneous vorticity for a MSW simulation with a mode-1 wave in $x$ and a mode-2 wave in $y$ of the same vorticity amplitude ($A = 0.5$), with respective frequencies 4.17 and 7.12. c) Lagrangian low-pass filter of the flow in panel a with a cut-off frequency of 2, so that both waves are removed, d) as in panel c with a cut-off frequency of 5.5, so that only the mode-2 wave is removed, and e) as in panel c with the filter defined in equation \eqref{lowbandpassfilter} and labelled `band-pass' in panel b, with $\omega_1 = 2$ and $\omega_2 = 5.5$, so that the mode-2 wave is retained and the mode-1 removed. f, g, and h) L2 wave perturbation corresponding to panels c, d, and e respectively. The directory including the Jupyter notebook that generated this figure can be accessed at \url{https://cocalc.com/share/public_paths/bdc0d1617e113644a25e3ba4c0b91b8fad20701f/Figure-7}.}
    \label{fig:lowbandpass}
\end{figure}

The Lagrangian wave definitions in figures \ref{fig:wave_decomp}i and \ref{fig:wave_decomp}l are much cleaner as they only represent the wave vorticity. However, the original mode-1 plane wave is not perfectly recovered, as can be seen in figures \ref{fig:wave_decomp}i and \ref{fig:wave_decomp}l. This is a result of nonlinear wave--mean interactions. There appear to be two such types of interaction -- large scale deformations of the plane wave due to interaction with the mean flow (seen more clearly in figure \ref{fig:lowbandpass} and supplementary movie 1), and high frequency oscillations of the mean flow that appear in figures \ref{fig:wave_decomp}i and \ref{fig:wave_decomp}l at the same spatial scales as the mean flow. The amplitude of this turbulence-like pattern scales with the wave linearity, is independent of grid resolution (making it unlikely to be caused by interpolation errors -- see \S \ref{sec:interpolation}), and is the same in both strategies 1 and 3. The time evolution of the mean and wave fields is shown in supplementary movie 1.

\begin{figure}
    \centering
    \includegraphics[width=\textwidth]{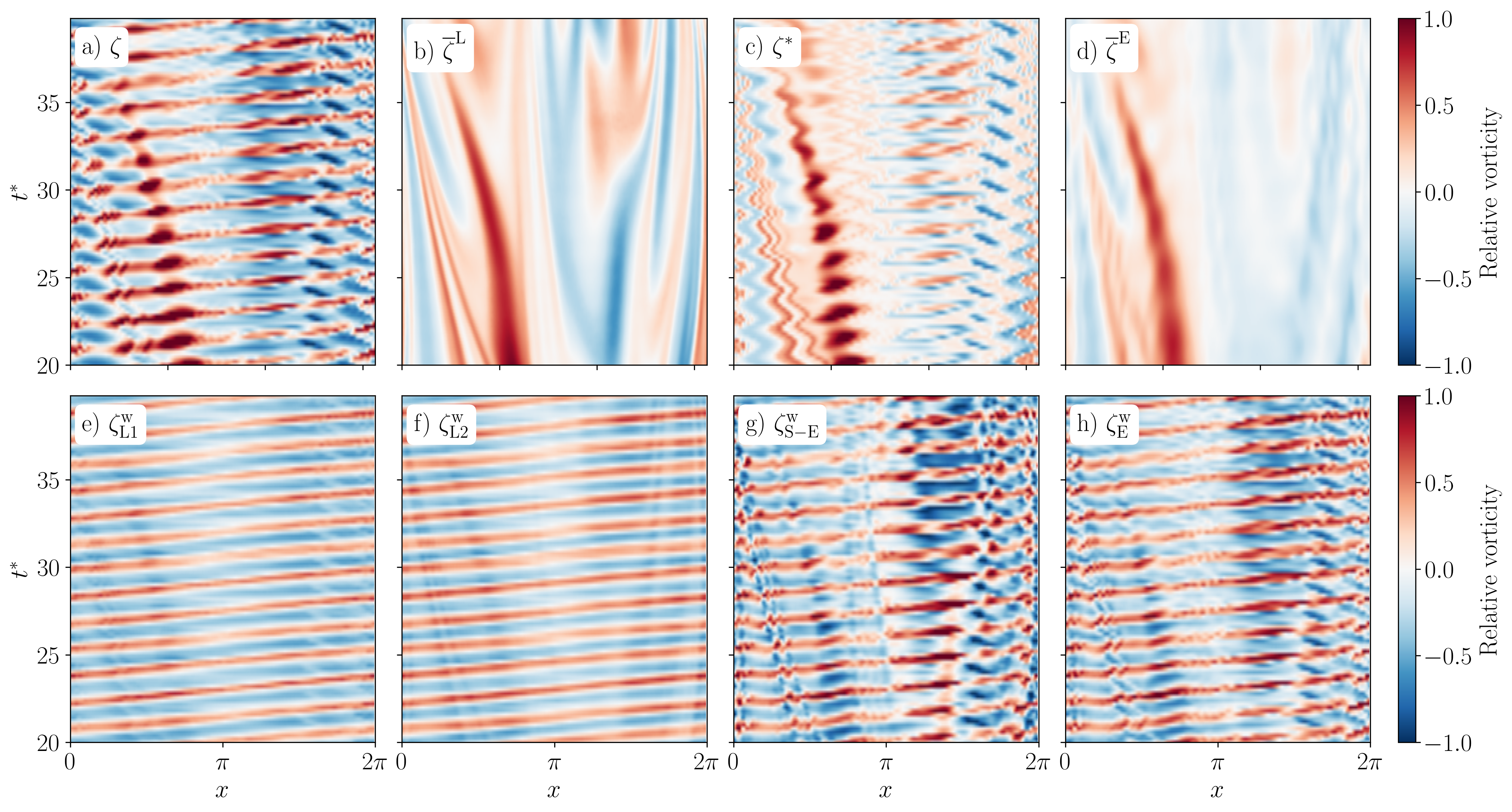}
    \caption{Hovm{\"o}ller (space-time) diagrams of vorticity: a) instantaneous, b) Lagrangian low-pass , c) Lagrangian low-pass  at the trajectory midpoint, d) Eulerian low-pass, e) Lagrangian L1 wave, f) Lagrangian L2 wave, g) semi-Eulerian wave, and h) Eulerian wave. Strategy 3 is used to solve for the Lagrangian means at a temporal resolution of 0.2. Parameters are identical to figure \ref{fig:mean_comparison}. All panels are shown at $y = 2.8$. The directory including the Jupyter notebook that generated this figure can be accessed at \url{https://cocalc.com/share/public_paths/bdc0d1617e113644a25e3ba4c0b91b8fad20701f/Figure-6}.}
    \label{fig:hovmoller}
\end{figure}

Figure \ref{fig:hovmoller} shows Hovm{\"o}ller diagrams of several of the fields shown in figure \ref{fig:wave_decomp}. Comparing the instantaneous (figure \ref{fig:hovmoller}a) and Lagrangian mean (figure \ref{fig:hovmoller}b) vorticity shows that the wave has been very effectively removed by the Lagrangian low-pass filter. 

The wave oscillations are very clear in $\zeta^*$ (figure \ref{fig:hovmoller}c) -- from which $\barL{\zeta}$ in figure \ref{fig:hovmoller}b is remapped. Wave oscillations are also visible in the Eulerian mean $\overline{\zeta}^{\,\mathrm{E}}$ in figure \ref{fig:hovmoller}d, and the vorticity gradients in the turbulent flow are overly smoothed, as shown in figure \ref{fig:mean_comparison}c. 

The various wave decompositions are shown in the bottom row of figure \ref{fig:hovmoller}, again demonstrating that the two Lagrangian definitions give a clean representation of the wave field, whereas the Eulerian and semi-Eulerian wave-definitions contain significant imprints of the turbulent flow. A movie showing the evolution of $\zeta$, $\zeta^*$, and $\barL{\zeta}$ over the time series shown in figure \ref{fig:hovmoller} is provided in the supplementary material (supplementary movie 1).

Despite the Lagrangian wave perturbations being visually `cleaner' in that they recreate more closely the plane wave with which the simulation was initialised, the physically appropriate wave definition for a given problem is likely to be context-dependent. 

Finally, we present an example to demonstrate the flexibility of Lagrangian frequency filtering. Figure \ref{fig:lowbandpass}a shows the instantaneous vorticity of a flow that has been initialised with a turbulent flow and mode-1 wave in the $x$ direction (as before), and also a mode-2 wave in the $y$ direction. The waves have the same amplitude in vorticity, and have frequencies 4.17 and 7.12 respectively. In figures \ref{fig:lowbandpass}c and \ref{fig:lowbandpass}f, the mean and L2 wave perturbations are shown for a low-pass filter at cut-off frequency $\omega_c = 2$, which removes both waves from the mean flow. Figures \ref{fig:lowbandpass}d and \ref{fig:lowbandpass}g are as for figures \ref{fig:lowbandpass}c and \ref{fig:lowbandpass}f with cut-off frequency $\omega_c = 5.5$, which retains the mode-1 wave in the `mean' flow (figure \ref{fig:lowbandpass}d), and leaves the mode-2 wave in the `wave' perturbation (figure \ref{fig:lowbandpass}g). In figures \ref{fig:lowbandpass}e and \ref{fig:lowbandpass}h the filter defined in \eqref{lowbandpassfilter} is used, such that the Lagrangian mean operation removes frequencies between $\omega_1 =2$ and $\omega_2 = 5.5$ and retains all other frequencies. Therefore, the mode-2 wave is kept as part of the `mean', and the mode-1 is in the `wave' perturbation. Each of the weight functions are shown in figure \ref{fig:lowbandpass}b. As in figure \ref{fig:wave_decomp}, large scale departures of the wave perturbations in figures \ref{fig:lowbandpass}g  and \ref{fig:lowbandpass}h from a perfect mode-2 and mode-1 plane wave respectively are again attributed to nonlinear interactions between the waves and between the waves and the turbulent flow. 

\section{Numerical errors and interpolation}\label{sec:interpolation}

There are two primary error sources in our method -- the truncation of the interval length and the interpolations, and these error sources are shared by particle tracking methods. Appendix \ref{app:intervaltimes} shows the impact of varying the half-interval length $T$. We found that increasing $T$ reduced the remaining wave-frequency oscillations, but increasing the interval length past $T = 20$ made negligible difference to the solutions, making $T=20$ our value of choice (so that $\omega_c T = 40$, and the averaging interval $2T$ is 12.7 times longer than the cut-off period). It may however be worth reducing $T$ and suffering a small error of this type to reduce the computational expense of the calculations. 

The other source of error comes from interpolation. In the methods discussed here, there are two types of interpolation. The first is performed at every time step of strategies 2 and 3 in the $RHS$ of equations \eqref{xi21eq}, \eqref{partial_f_evol2}, \eqref{partial_f_evol3} and \eqref{xi31eq}, and requires finding a scalar field (the scalar to be averaged, or each component of velocity) at some coordinate $\vXip{2\mapsto 1}(x,t)$ (strategy 2) or $\vXip{3\mapsto 1}(x,t)$ (strategy 3), where the scalar is known on a regular grid. The second is the final remapping of $\tilde{f}$ (in strategy 1) or $f^*$ (in strategy 3) to $\barL{f}$. In this case, $\barL{f}$ is known at the irregularly spaced locations $\vXia{1\mapsto2}(\vx,t^*)$ or $\vXia{3\mapsto2}(\vx,t^*)$, and needs to be found at regularly gridded locations. We do not expect this second type of interpolation to be problematic in strategy 3, as demonstrated in figure \ref{fig:strat13_comparison}. 

When the flow is such that distances between points advected by the flow become very far apart or very close together over the interval of interest, as represented by an interpolating map with sharp gradients (e.g. $\vXia{1\mapsto 2}$ shown in figure \ref{fig:strat13_comparison}c), both types of interpolation can be prone to error. This can be the case when the flow is compressible \citep[or equivalently when performed on a 2D surface in a 3D incompressible flow;][]{Shakespeare2021a}, but also when the flow is incompressible and straining or shearing. 

In the shallow water case here, the flow is both compressible and straining/shearing, and the mean flow evolves significantly over the time interval (e.g. figure \ref{fig:mean_comparison_hovmuller}b). We therefore expect to accumulate the first type of interpolation error at each time step, although we do not see evidence of this error in our experiments. The mean flow `imprint' in the wave component in figures \ref{fig:wave_decomp}i, \ref{fig:wave_decomp}l, \ref{fig:hovmoller}c, and  \ref{fig:hovmoller}f is independent of resolution and scales with wave nonlinearity, so we attribute this to physical nonlinear interactions as discussed in \S \ref{sec:waveresults}. 

Although strategies 2 and 3 accumulate interpolation error at every time step, the interpolation terms in the Lagrangian scalar equations \eqref{partial_f_evol2} and \eqref{partial_f_evol3} are weighted by $G(t^* - t)$, which becomes small as the time moves away from the interval midpoint (see figure \ref{fig:mean_comparison}d). Therefore, the (more accurate) interpolation along trajectories over short times is more important, and long time interpolations become negligible. Increasing the interval time to reduce truncation error does not significantly increase interpolation error in strategy 3. 

Strategy 1 does not suffer from accumulation of interpolation error as particle tracking and the other strategies do, needing only one interpolation, but this interpolation can be complex and inaccurate (as shown in figure \ref{fig:strat13_comparison}), and depends strongly on the length of the averaging interval. Thus increasing the averaging interval worsens interpolation error in strategy 1, but improves truncation error.

A choice over whether strategy 1,2 or 3 is optimal should be made based on the nature of the flow, time-scales, boundary conditions, computational parallelisation, and weight function involved. 

\section{Discussion}\label{sec:discussion}

In this work, we have extended the PDE-based approach of KV23 for finding a top-hat Lagrangian mean to a Lagrangian mean with a general convolutional weight function. In particular, this has allowed us to present a method for Lagrangian frequency filtering, whereby specific intrinsic frequencies of a flow can be isolated from the rest of the flow. We have also derived some of the special properties of Lagrangian mean flows that hold for particular weight functions, and explored several different wave--mean decompositions.

In addition to re-deriving the strategies 1 and 2 of KV23 for a general weight function, we have presented a novel strategy 3 that removes some of the difficulties associated with strategies 1 and 2, and have shown that this strategy allows a clean decomposition of geostrophic turbulence and large amplitude Poincar{\'e} waves in a simple rotating shallow water system. 

In the system presented here, Lagrangian filtering aims to recover the mean flow without the signature of the large amplitude wave displacements. We have demonstrated the ability of our method to achieve this. However, an equally important use of Lagrangian filtering is to allow decomposition of waves and mean flows when the waves have been Doppler shifted by the mean flow, that is when the flow speed is large compared to the phase speed of the waves. This is the use that \citet{Shakespeare2021a} focus on in their presentation of Lagrangian filtering. Although we do not show an example of Doppler shifting of waves by the mean flow here, the method is straightforwardly applicable.

Our method for Lagrangian filtering can be compared to existing particle tracking methods. Lagrangian filtered fields can be found by tracking particles online, although Lagrangian means are then defined at and remapped from the initially seeded particle positions. This can lead to problems with particle clustering similar to those discussed in \S \ref{sec:interpolation}. To tackle this problem, \citet{Shakespeare2021a} recently presented an open-source implementation of offline Lagrangian filtering. Their method uses offline simulation data (scalars and horizontal velocities) to track particles backwards and forwards from the interval midpoint, finding the time series of a scalar on a particle, temporally filtering, and assigning the filtered scalar to the trajectory midpoint. This method (when carried out using 3D velocities) directly finds $f^*$ (in our notation), rather than the generalised Lagrangian mean $\barL{f}$, but this may be sufficient if the waves are low amplitude and $f^* \simeq \barL{f}$, or if the $\mathrm{L1}$ wave decomposition is needed. Alternatively, $\barL{f}$ could be recovered by finding the Lagrangian mean of position and performing an interpolation of $f^*$ to $\barL{f}$ (similar to the final step in our strategy 3). Offline particle tracking requires saving, storing, and processing large quantities of simulation output, requiring high storage and post-processing cost. Particle tracking also suffers from expense and error associated with interpolation at each time step, similarly to our strategies 2 and 3. 

In contrast, our method solves the Lagrangian mean equations at the same time as the evolution equations of the flow itself, so saving high frequency simulation output is not required. This also allows the Lagrangian equations to be solved on the same grid and using the same numerical scheme as the original simulation. There is flexibility over the weight function used and the specific Lagrangian mean and wave definitions that are solved for. However, this does increase the computational expense of the simulation itself. In our 2D shallow water example, using (the cheapest) strategy 1 to solve for $\barL{f}$ increased the computation time over shallow water alone by 66\%, and using strategy 3 by 132\% (see Appendix \ref{app:expense}). 

We therefore expect the two different methods to have different uses. When filtering existing simulation output, or output from a large and complex general circulation model, it may be preferable to use particle tracking offline. However, for process studies where wave or mean identification is a primary objective, our method is easily implemented, more flexible, and requires much less storage. 

When finding the Lagrangian mean at high temporal resolution, the expense of our method increases greatly since a set of Lagrangian mean equations needs to be solved for each time $t^*$ where the Lagrangian mean is required (e.g. as in figure \ref{fig:hovmoller}). Particle tracking methods also suffer from this drawback to some extent. If a `slow' Lagrangian mean is the quantity of interest, then a time series of Lagrangian means can be found at a coarse time resolution that only resolves this slow variation. The semi-Eulerian wave perturbation can then be found at all times by interpolating this mean to the time of interest and removing it from the instantaneous field. However, if one of the Lagrangian wave definitions is needed, then the Lagrangian mean calculation needs to be carried out at a temporal resolution that captures the waves. 

There is however a special class of weight functions that are exponential or sum-of-exponentials, in which case the partial Lagrangian mean found at each time step during the evolution of the Lagrangian mean equations is itself the full mean \citep{minzExponentialLagrangianMean2024}. This allows the Lagrangian mean to be found at each time step with the expense of solving only one set of Lagrangian mean equations. The drawback of this method is an inability to freely choose the weight function (for example, to filter at a specific frequency) but the significant improvement in computational cost may make this worthwhile. The derivation and an evaluation of the exponential mean is presented in \citet{minzExponentialLagrangianMean2024}. 

We presented three separate strategies for finding the Lagrangian mean, each of which has its own advantages and disadvantages. Strategy 1 is cheapest and is simple to implement (particularly in Distributed Memory Parallelisation), but fails when the mean flow varies significantly over the averaging interval (e.g. figure \ref{fig:strat13_comparison}b). Strategy 2 directly finds the Lagrangian mean, but cannot easily be used in bounded domains and is the slowest method (see Appendix \ref{app:expense}). 

Strategy 3 is slower than strategy 1, but faster than strategy 2. It has simple boundary conditions, and can be solved in a periodic or bounded domain. Like strategy 1, it requires a final interpolation step, but this interpolation is simpler than that in strategy 1, and likely to be accurate for a low-pass filter. Future work will focus on implementing the three strategies in a 3D Boussinesq solver and testing their ability for Lagrangian filtering of different flow configurations. 

\backsection[Supplementary data]{A supplementary movie is provided (supplementary movie 1).} 

\backsection[Acknowledgements]{We are grateful to the editor and to two anonymous reviewers for their helpful and constructive comments.}

\backsection[Funding]{L.E.B. is supported by the Engineering and Physical Sciences Research Council, grant EP/X028135/1. H.A.K. is supported by by the Engineering and Physical Sciences Research Council, grant EP/Y021479/1. J.V. is supported by the UK Natural Environment Research Council, grant NE/W002876/1.}

\backsection[Declaration of Interests]{The authors report no conflict of interest.}

\backsection[Data availability statement]{The data that support the findings of this study are openly available at  https://doi.org/10.5281/zenodo.14237745 \citep{bakerSWGLMsoftware24}. }
\backsection[Author ORCIDs]{L.E. Baker, https://orcid.org/0000-0003-2678-3691; H. A. Kafiabad, https://orcid.org/0000-0002-8791-9217; C. Maitland-Davies, https://orcid.org/0000-0001-9195-2477; J. Vanneste, https://orcid.org/0000-0002-0319-589X}

\appendix
\section{The mean of a Lagrangian mean scalar}\label{app:meanofmean}

As explained in \S \ref{sec:definingmean}, we would like the mean $\barL{f}(\vx,t^*)$ to satisfy a property that we expect of a mean, namely that the mean is unchanged by reapplying the averaging operation \eqref{meanofmean}.

Here, we introduce some clarifying notation to define what is meant by the operation on the $LHS$ of \eqref{meanofmean}, since the averaging operation itself depends on the flow with respect to which the Lagrangian mean is taken. We define Lagrangian averaging operators that act on some scalar field $h(\vx,t)$:
\begin{align}
    \overline{\,h\,}^{\,\vp}(\va,t^*)  &= \int_{-\infty}^{\infty} G(t^*-s)h(\vp(\va,s),s) \,\d s \label{phiaveragedef}\\
    \overline{\,h\,}^{\,\vpbar}(\va,t^*) &= \int_{-\infty}^{\infty} G(t^*-s)h(\vpbar(\va,s),s) \,\d s\,,\label{phibaraveragedef}
\end{align}
such that $\overline{\,(\cdot)\,}^{\,\vp}$ denotes a Lagrangian mean at time $t^*$ along the flow defined by $\vp$ for a trajectory labelled by $\va$, and $\overline{\,(\cdot)\,}^{\,\vpbar}$ similar for a (mean) flow defined by $\vpbar$. The definition of the generalised Lagrangian mean $\barL{f}$ of a scalar $f$ given in \eqref{fbardef} can then be written:
\begin{equation}\label{compositiondef}
    \barL{f}(\vpbar(\va,t^*),t^*) \equiv \left(\barL{f}\circ \vpbar\right) (\va,t^*) \equiv \overline{\,f\,}^{\,\vp}(\va,t^*)\,,
\end{equation}
where function composition is denoted by `$\circ$' and taken to apply to the first argument of a given function. Note that $\overline{\,f\,}^{\,\vp}$ is a function of the label space $(\va,t^*)$, making it a fully Lagrangian variable, whereas $\barL{f}$ is a function of physical space (specifically, the mean position).

The Lagrangian mean of the Lagrangian mean scalar should be taken with respect to the mean flow, so \eqref{meanofmean} can now be posed more carefully as
\begin{equation}\label{meanofmean2}
    \overline{\,\barL{f}\,}^{\,\vpbar} = \overline{\,f\,}^{\,\vp}\,.
\end{equation}
We find the conditions on the weight function $G$ for which \eqref{meanofmean2} holds. We have
\begin{align}
\overline{\,\barL{f}\,}^{\,\vpbar}(\va,t^*) &= \int_{-\infty}^{\infty} G(t^* -s)\barL{f}(\vpbar(\va,s),s)\,\d s \\
    &= \int_{-\infty}^{\infty}\left[\int_{-\infty}^{\infty} G(t^* - u)G(u-s)\,\d u\right] f(\vp(\va,s),s)\,\d s\,.
\end{align}
By comparison with \eqref{fbardef}, we see that
\begin{align}
    \overline{\,\barL{f}\,}^{\vpbar}(\va,t^*) = \overline{\,f\,}^{\,\vp}(\va,t^*) &\Leftrightarrow \int_{-\infty}^{\infty} G(t^* - u)G(u-s)\,\d u = G(t^*-s)\label{freq_condition1}\\
    &\Leftrightarrow \left(\hat{G}(\omega)\right)^2 = \hat{G}(\omega)\,,\label{freq_condition}
\end{align}
where $\hat{G}(\omega)$ is the Fourier transform of $G$, defined in \eqref{FTF}. 

From \eqref{freq_condition}, we see that condition \eqref{meanofmean2} is only satisfied if $\hat{G}(\omega) = 0$ or $1$, or some piece-wise combination of each.

\section{Relation to classical GLM theory and time-scale separation}\label{app:timescales}

The development of GLM theory by \citet{andrewsExactTheoryNonlinear1978} defines averaging procedures in an abstract way that apply similarly to spatial, temporal, or ensemble Lagrangian averages. However, for a time average it is assumed that there is a time-scale separation between the `slow' and `fast' motions to be separated by Lagrangian averaging \citep{Buhler2014book}. Here, we explain this assumption and  how it relates to our formulation.

First, we define the lift map $\vXi$ from the mean flow map to the flow map (this is equivalent to $\vXia{2 \mapsto 3}$ in the notation of the current work) by
\begin{equation}\label{liftmapdef}
    \vXi(\vpbar(\va,t^*),t^*) = \vp(\va,t^*)\,,
\end{equation}
so that a particle at position $\vXi(\vx,t)$ has mean position $\vx$ at time $t$. In our notation (using the definitions \eqref{phiaveragedef}-\eqref{phibaraveragedef}), we have
\begin{equation}\label{flowmaprelations}
    \overline{\,f \circ \vXi\,}^{\,\vpbar} = \overline{\,f\,}^{\,\vp}\,.
\end{equation}
\citet{andrewsExactTheoryNonlinear1978} define the Lagrangian mean (which we distinguish from our definition by a prime) as
\begin{equation}\label{AMLMdef}
    \overline{f}^{\mathrm{L}'}(\vx,t^*) = \overline{\,f \circ \vXi\,}^{\,\mathrm{E}}(\vx,t^*)\,,
\end{equation}
where $\overline{\,(\cdot)\,}^{\,\mathrm{E}}$ denotes the Eulerian time average defined in \eqref{fbarEdef}.

Rewriting our definition of the Lagrangian mean for comparison with \eqref{AMLMdef} using \eqref{compositiondef} and \eqref{flowmaprelations} gives
\begin{equation}
    \barL{f} \circ \vpbar = \overline{\,f \circ \vXi\,}^{\,\vpbar}\,.
\end{equation}
Therefore, the definitions of $\barL{f}$ and $\overline{f}^{\mathrm{L}'}$ are equivalent (and the assumption of separate slow and fast time-scales holds) when
\begin{equation}
    \overline{\,f \circ \vXi\,}^\mathrm{E}\circ \vpbar = \overline{\,f \circ \vXi\,}^{\vpbar}\,,
\end{equation}
or equivalently, when the Eulerian mean $\overline{\,(\cdot)\,}^{\,\mathrm{E}}$ in the definition \eqref{AMLMdef} is a good approximation to the more general expression $\overline{\,(\cdot)\,}^{\,\vpbar}\circ \vpbar^{-1}$\,.

To demonstrate this condition in a different way, we can write
\begin{align}
\overline{f}^{\mathrm{L}'}(\vx,t^*) &= \overline{\,f \circ \vXi\,}^{\,\mathrm{E}}(\vx,t^*) \\
&= \int_{-\infty}^{\infty} G(t^*-s) f(\vXi(\vx,s),s) \,\d s\,.
\end{align}
Letting $\vx = \vpbar(\va,t^*)$ gives
\begin{equation}\label{AMmean}
\overline{f}^{\mathrm{L}'}(\vpbar(\va,t^*),t^*) = \int_{-\infty}^{\infty} G(t^*-s)f(\vXi(\vpbar(\va,t^*),s),s)\,\d s\,.
\end{equation}
However, by our definitions \eqref{phiaveragedef} and \eqref{phibaraveragedef},
\begin{align}
\barL{f}(\vpbar(\va,t^*),t^*)&= \int_{-\infty}^{\infty} G(t^*-s)f(\vp(\va,s),s),s)\,\d s \\
&= \int_{-\infty}^{\infty} G(t^*-s)f(\vXi(\vpbar(\va,s),s),s)\,\d s \, .\label{ourmean}
\end{align}
The two definitions $\barL{f}$ and $\overline{f}^{\mathrm{L}'}$ are approximately equal provided that (comparing \eqref{AMmean} and \eqref{ourmean}) $f(\vXi(\vpbar(\va,t^*),s),s) \approx f(\vXi(\vpbar(\va,s),s),s)$ where $G(t^*-s)$ is not small. This assumes that the mean flow is `frozen' during the averaging operation and is the implicit assumption of time-scale separation behind the \citet{andrewsExactTheoryNonlinear1978} definition of the Lagrangian mean. 

A flow which illustrates the difference between the two formulations is the classical lee-wave problem discussed in \S \ref{sec:wavemeandecomp}. Consider the flow defined in \eqref{beq} - \eqref{ueq}. The flow is steady, so $\vXi$ (as defined in \eqref{liftmapdef}) and the scalar $f$ to be averaged are independent of time. Then we have
\begin{align}
\overline{f}^{\mathrm{L}'}(\vx) &= \int_{-\infty}^{\infty} G(t^*-s) f(\vXi(\vx)) \,\d s \label{leewave_AM_mean1}\\
&= f(\vXi(\vx))\,,\label{leewave_AM_mean2}
\end{align}
whereas
\begin{equation}
\barL{f}(\vpbar(\va,t^*)) = \int_{-\infty}^{\infty} G(t^*-s)f(\vp(\va,s))\,\d s\,.
\end{equation}
Suppose the scalar $f$ is the vertical velocity $w$. The first definition (\eqref{leewave_AM_mean1} - \eqref{leewave_AM_mean2}) is simply a rearrangement of the scalar field and no averaging is performed, so the Lagrangian mean will be non-zero. However, the second definition (as used in this paper) will average over the oscillations of $w$ on a particle and give zero Lagrangian mean (when $G$ is defined to remove wave frequencies) as is expected. The difference between the two definitions results from relaxation of the assumption in the second that the mean flow is `frozen' during the averaging interval.

\section{The Lagrangian material derivative}\label{app:materialderivative}

We show that, for convolutional weight functions $G(t^*-s)$, there is a powerful relation between the material derivative of a Lagrangian mean quantity and the Lagrangian mean of a material derivative, namely (c.f. \eqref{materialderivativerel}--\eqref{materialderivativedef}, repeated here)
\begin{equation}\label{materialderivativerel2}
    \bar{\bar{D}}\barL{f} = \overline {\,D f\,}^{\mathrm{L}}\,,
\end{equation}
where
\begin{align}\label{materialderivativedef2}
    D &\equiv \frac{\partial}{\partial t} + \vu\bcdot \nabla \,, \\
    \bar{\bar{D}} &\equiv \frac{\partial}{\partial t^*} + \bar{\bar{\vu}}\bcdot \nabla \,.
\end{align}

We have
\begin{align}
    \left(\bar{\bar{D}}\barL{f}\right) \circ \vpbar(\va,t^*) &= \left( \frac{\partial \barL{f}}{\partial t^*} + \bar{\bar{\vu}}\bcdot \nabla \barL{f} \right) \circ \vpbar(\va,t^*)\label{LMD1} \\
    &= \frac{\d}{\d t^*}\left(\barL{f}(\vpbar(\va,t^*),t^*)\right) \label{LMD2}\\
    &= \frac{\d}{\d t^*}\int_{-\infty}^\infty G(t^*-s)f(\vp(\va,s),s)\,\d s \label{LMD3}\\
    &= \int_{-\infty}^\infty G'(t^*-s) f(\vp(\va,s),s)\,\d s\,\label{LMD4}\\
    &= \int_{-\infty}^\infty G(t^*-s) \frac{\d}{\d s}\left(f(\vp(\va,s),s)\right) \, \d s \label{LMD5}\\
    &= \int_{-\infty}^\infty G(t^*-s)\left(\frac{\partial f}{\partial s}(\vp(\va,s),s) + \frac{\partial \vp}{\partial s}(\va,s) \bcdot \nabla f(\vp(\va,s),s)\right)\, \d s\label{LMD6} \\
    & = \left(\overline{\,D f\,}^{\mathrm{L}}\right)\circ \vpbar(\va,t^*)\,,
\end{align}
where from \eqref{LMD1}-\eqref{LMD2} we used the definition \eqref{LMUdef} of Lagrangian mean velocity, and from \eqref{LMD4}-\eqref{LMD5} we relied on the convolutional form of the weight function and used integration by parts, assuming from \eqref{normalisation} that $G(t^*-s) \rightarrow 0$ as $s\rightarrow \pm \infty$. Equivalently, it can be shown that \eqref{materialderivativerel2} only holds when the weight function takes the specific (convolutional) form of a frequency filter. 

\section{Derivation of strategy 3}\label{app:strat3}
Here we derive strategy 3, which solves directly for $f^*(\vp(\va,t^*),t^*)$. For this we need to solve for the map $\vXia{3\mapsto 1}$, and also find a map $\vXia{3\mapsto 2}$ to enable us to find $\barL{f}(\vpbar(\va,t^*),t^*)$. 

We first consider the case $t < t^*$. This case is equivalent to strategy 1, since we solve at $\vpstar(\va,t) = \vp(\va,t)$. Differentiating \eqref{fbarpdefs}, or by comparison with \eqref{partial_f_evol1} in strategy 1, we can write an equation for the partial mean $f^*_p$:
\begin{equation}
    \frac{\partial f^*_p}{\partial t}(\vx,t) + \vu(\vx,t)\bcdot \nabla f^*_p(\vx,t) = f(\vx,t)G(t^* - t)\,. \label{partial_f_evol31}
\end{equation}
We note that for $t < t^*$, $\vXip{3\mapsto 1}(\vx,t) = \vx$ is the identity map, so, defining 
\begin{equation}
\vxip{3\mapsto 1}(\vx,t) = \vXip{3\mapsto 1}(\vx,t) - \vx\,,
\end{equation}
we have
\begin{equation}\label{xi31zero}
    \vxip{3\mapsto 1}(\vx,t) = 0\,, \hspace{1cm} t<t^*\,.
\end{equation}
We also solve for $\vXia{3\mapsto 2}(\vx,t^*)$ to enable us to find $\barL{f}(\vx,t^*)$. To do this, we differentiate the definition of $\vXip{3\mapsto 2}$ in \eqref{Xipartialmaps3}. For $t < t^*$, this is equivalent to solving for $\vXip{1\mapsto 2}$ in strategy 1, so we define (c.f. \eqref{xiXi1})
\begin{equation}\label{Xixi3}
    \vxip{3\mapsto 2}(\vx,t) = \vXip{3\mapsto 2}(\vx,t) - \vx\,,
\end{equation}
then from \eqref{xi12eq},
\begin{equation}\label{xi32eq1}
    \frac{\partial \vxip{3\mapsto 2}}{\partial t}(\vx,t) + \vu(\vx,t)\bcdot \nabla \vxip{3\mapsto 2} (\vx,t) = - \vu(\vx,t)\int_{t^*-T}^t G(t^*-s)\,\d s\,.
\end{equation}
We now consider the case $t > t^*$. Differentiating \eqref{fbarpdefs} with respect to $t$, we find
\begin{equation}\label{partial_f_evol32}
    \frac{\partial f^*_p}{\partial t}(\vx,t) = G(t^* - t)f(\vx + \vxip{3\mapsto 1}(\vx,t))\,.
\end{equation}
We now find an equation for $\vxip{3\mapsto 1}(\vx,t)$. Differentiating the definition of $\vXip{3\mapsto 1}$ in \eqref{Xipartialmaps2} gives
\begin{align}
    \frac{\partial \vXip{3\mapsto 1}}{\partial t} &= \vu(\vXip{3\mapsto 1}(\vx,t),t)\\
    \Leftrightarrow \hspace{1cm} \frac{\partial \vxip{3\mapsto 1}}{\partial t} &= \vu(\vx + \vxip{3\mapsto 1}(\vx,t),t)\,.\label{xi31eq2}
\end{align}
Finally, we find an equation for $\vxip{3\mapsto 2}(\vx,t)$. Differentiating the definition of $\vXip{3\mapsto 2}$ in \eqref{Xipartialmaps3} gives
\begin{align}
    \frac{\partial \vXip{3\mapsto 2}}{\partial t}(\vpstar(\va,t),t) &= \vu(\vp(\va,t),t)\left(1 - \int_{t^*-T}^t G(t^*-s)\,\d s\right)\\
    \Leftrightarrow \hspace{1cm} \frac{\partial \vxip{3\mapsto 2}}{\partial t}(\vx,t) &= \vu(\vx + \vxip{3\mapsto1}(\vx,t),t)\left(1 - \int_{t^*-T}^t G(t^*-s)\,\d s\right)\,.\label{xi32eq2}
\end{align}
The final equations are summarised as \eqref{partial_f_evol3}--\eqref{xi32eq} in the main text.

\section{Run time of each strategy}\label{app:expense}
Table \ref{tab:expense} shows the run time of strategies 1,2, and 3 when solving for different combinations of the Lagrangian fields that may be required. Simulations are run at $256 \times 256$ horizontal resolution over 20 time units ($T=10$) with a time step of 0.005 (4000 time steps), and an average time taken over three runs. The time reported is for the simulation only -- the final remapping in each case takes the same time as $< 100$ time steps. Vorticity is the only Lagrangian mean scalar being solved for. 

The variation in run times between each column of table \ref{tab:expense} for each strategy is due to the number of equations being solved and the complexity of these equations. The combinations of PDEs needed for each strategy are shown in table \ref{tab:strat}. 

For each combination of Lagrangian fields to be solved, strategy 1 is the fastest because there is no interpolation on the $RHS$ of the Lagrangian mean equations \eqref{partial_f_evol1},\eqref{xi12eq}, and \eqref{xi13eq}. However, this can come at the expense of accuracy due to the final interpolation needed to recover $\barL{f}$ (see figure \ref{fig:strat13_comparison}).

The most expensive operations in this pseudo-spectral solver are finding interpolations and calculating nonlinear terms. Strategy 3 is faster than strategy 2 because strategy 2 requires finding both a nonlinear advection term \textit{and} an interpolated term at every time step in equations \eqref{xi21eq}, \eqref{partial_f_evol2}, and \eqref{xi23eq}, whereas strategy 3 (equations \eqref{partial_f_evol3}, \eqref{xi31eq}, and \eqref{xi32eq}) requires \textit{either} computing nonlinear terms \textit{or} computing interpolations at each time step, not both (since $\vx + \vxip{3\mapsto1}(\vx,t) = \vx$ for $t < t^*$ from \eqref{xi31zero}).

\begin{table}
  \begin{center}
\def~{\hphantom{0}}
  \begin{tabular}{lccc}\label{expense_table}
       \bf{Strategy} & \bf{Solve for $\barL{f}$}\hspace{0.5cm}   &   \bf{Solve for $f^*$}\hspace{0.5cm} & \bf{Solve for $f^*$ and $\barL{f}$} \\[5pt]
       Strategy 1 & 1 & 1.01 & 1.19\\[3pt]
       Strategy 2 & 2.13 & 2.31 & 2.31\\[3pt]
       Strategy 3 & 1.39 & 1.30 & 1.39\\[3pt]
       
  \end{tabular}
  \caption{Run times for the shallow water simulation and Lagrangian mean computation for each strategy, using the code given in \cite{bakerSWGLMsoftware24}. Times are normalised by the time taken for strategy 1 when solving for $\barL{f}$ only. For comparison, when the simulation is run without the Lagrangian mean equations (shallow water only), the corresponding normalised time is 0.6.}
  \label{tab:expense} 
  \end{center}
\end{table}

\section{Comparison of filter interval times}\label{app:intervaltimes}

Throughout this study, we used an averaging interval time of $2T = 40$. Longer averaging times improve the wave decomposition, since there is less truncation error when approximating the full Lagrangian mean \eqref{fbardef} by the integral over the finite interval \eqref{fbardefs}. The condition for the truncation to approximate the full interval is $\omega_c T \gg 1$, where $\omega_c = 2$ here.

\begin{figure}
    \centering
    \includegraphics[width=\textwidth]{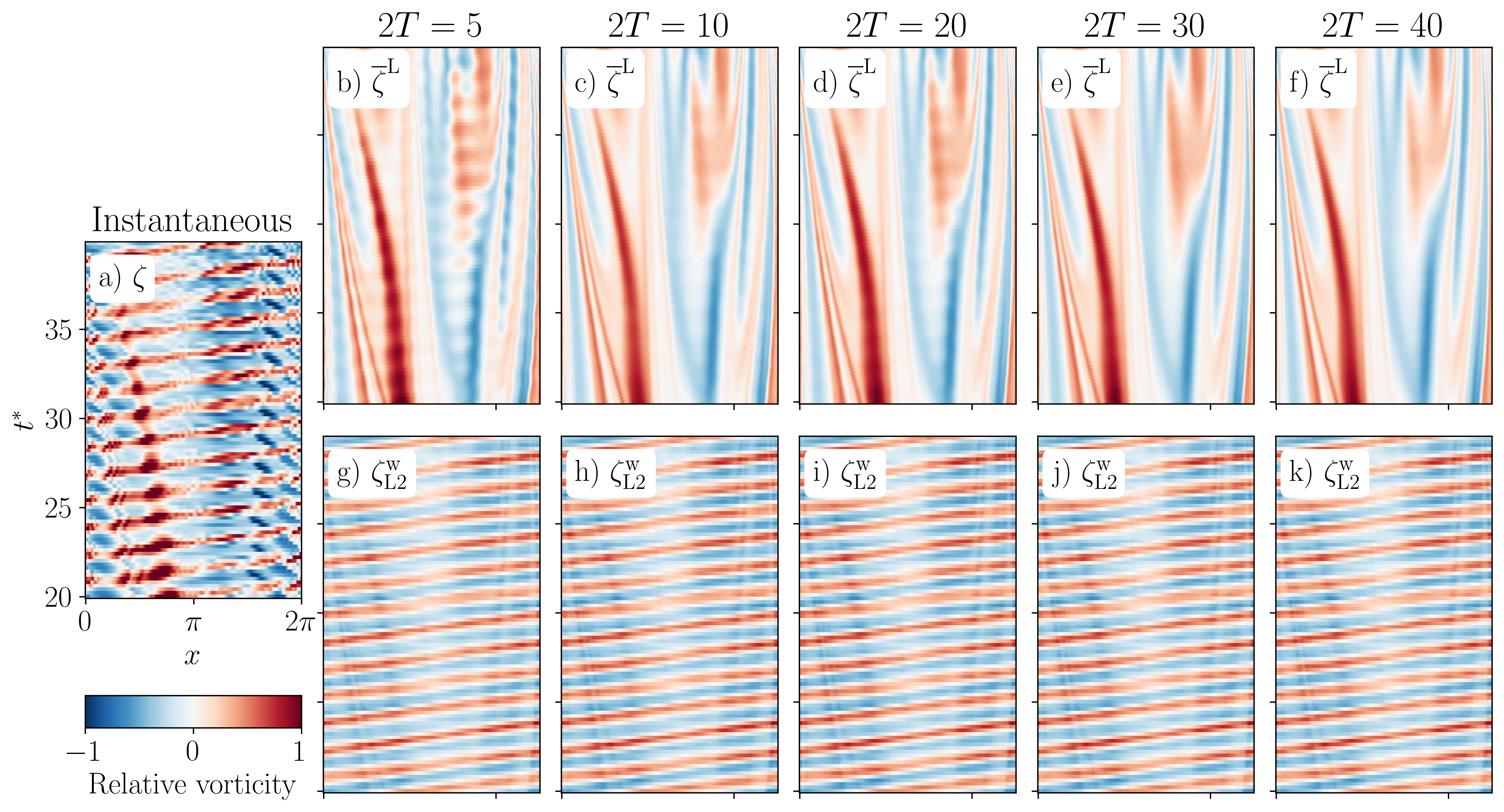}
    \caption{Hovm{\"o}ller diagrams of vorticity for a range of averaging interval times. a) instantaneous $\zeta$, (top row) Lagrangian mean $\barL{\zeta}$, and (bottom row) L2 wave $\zeta_{\mathrm{L2}}^\mathrm{w}$. The directory including the Jupyter notebook that generated this figure can be accessed at \url{https://cocalc.com/share/public_paths/bdc0d1617e113644a25e3ba4c0b91b8fad20701f/Figure-8}.}
    \label{fig:interval_comparison}
\end{figure}

Figure \ref{fig:interval_comparison} shows the impact of increasing the interval time $2T$on the time series of Lagrangian mean and L2 wave perturbation. As $T$ increases, the quality of the filter improves and progressively more wave signal is removed from the Lagrangian mean. The error decreases until $2T = 40$. Filters that are more localised in time (such as a Butterworth or Gaussian filter) would also allow earlier truncation and a shorter averaging interval. 

\bibliographystyle{jfm}
\bibliography{Library}

\end{document}